\documentclass[lineno]{jfm_arxiv}
\usepackage{graphicx,subcaption,hyperref}
\usepackage{mathtools}
\usepackage{newtxtext}
\usepackage{newtxmath}
\usepackage{natbib}
\usepackage{hyperref}
\usepackage{xcolor}
\usepackage{cases}
\usepackage{overpic}
\usepackage{siunitx}
\hypersetup{
    colorlinks = true,
    urlcolor   = blue,
    citecolor  = black,
}

\newcommand{\RomanNumeralCaps}[1]
\linenumbers

\newcommand\Wi{\mathit{Wi}}
\newcommand\Sc{\mathit{Sc}}
\newcommand{\trC}{\operatorname{tr}\mathsfbi{C}}
\newcommand{\detC}{\det\mathsfbi{C}}

\title{Preserving large-scale features in simulations of elastic turbulence}
\author{Sumithra R. Yerasi\aff{1},
Jason R. Picardo\aff{2}
  \corresp{\email{picardo@iitb.ac.in}; 
  },
  Anupam Gupta\aff{3},
 \and Dario Vincenzi\aff{1}
 }

\affiliation{\aff{1}Universit\'e C\^ote d'Azur, CNRS, LJAD, 06100 Nice, France
\aff{2}Department of Chemical Engineering, Indian Institute of Technology Bombay, Mumbai 400076, India
\aff{3}Department of Physics, Indian Institute of Technology Hyderabad, Hyderabad 502284, India
}

\begin{document}
\maketitle

\begin{abstract}

Simulations of elastic turbulence, the chaotic flow of highly elastic and inertialess polymer solutions, are plagued by numerical difficulties: The chaotically advected polymer conformation tensor develops extremely large gradients and can loose its positive definiteness, which triggers numerical instabilities. While efforts to tackle these issues have produced a plethora of specialized techniques---tensor decompositions, artificial diffusion, and shock-capturing advection schemes---we still lack an unambiguous route to accurate and efficient simulations. In this work, we show that even when a simulation is numerically stable, maintaining positive-definiteness and displaying the expected chaotic fluctuations, it can still suffer from errors significant enough to distort the large-scale dynamics and flow-structures. We focus on two-dimensional simulations of the Oldroyd-B and FENE-P equations, driven by a large-scale cellular body-forcing. We first compare two positivity-preserving decompositions of the conformation tensor: symmetric square root (SSR) and Cholesky with a logarithmic transformation (Cholesky-log). While both simulations yield chaotic flows, only the latter preserves the pattern of the forcing, \textit{i.e.}, its fluctuating vortical cells remain ordered in a lattice. In contrast, the SSR simulation exhibits distorted vortical cells which shrink, expand and reorient constantly. To identify the accurate simulation, we appeal to a hitherto overlooked mathematical bound on the determinant of the conformation tensor, which unequivocally rejects the SSR simulation. Importantly, the accuracy of the Cholesky-log simulation is shown to arise from the logarithmic transformation. We then consider local artificial diffusion, a potential low-cost alternative to high-order advection schemes. 
Unfortunately, the diffusive smearing of polymer stress in regions of intense stretching is found to destabilize the adjacent vortices, thereby modifying the global dynamics. We end with an example, showing how the spurious large-scale motions, identified here, contaminate predictions of scalar mixing.
\end{abstract}



\section{Introduction}
\label{sec:intro}

At low inertia but high elasticity, polymer solutions develop a chaotic regime called elastic turbulence \citep{gs00}.   
As the name suggests, in this regime the velocity field is chaotic and fluctuates in space and time with power-law spectra. Unlike Newtonian turbulence, though, the instabilities that trigger the chaotic motions and the mechanisms that sustain them are solely elastic and fluid inertia plays no role. 
Consequently, elastic turbulence offers an effective means of enhancing mixing under circumstances of weak or vanishing inertia, such as in micro-fluidics \citep{gs01}. 
The discovery of this phenomenon raised new questions, most of which remain unanswered, including the role of flow geometry and perturbations in triggering and sustaining turbulence, and the significance of wave-like fluctuations in the spatio-temporal dynamics~\citep{datta22}. Therefore, over the past two decades, elastic turbulence has been the subject of extensive experimental investigation \citep{steinberg21,steinberg22}. Numerical simulations, however, have lagged behind and to date only a few studies have considered realistic three-dimensional flows \citep{liu13,stark22,song22,song23}. Indeed, simulating elastic turbulence presents a distinct set of difficulties and in some ways is more challenging than simulating its Newtonian counterpart \citep*{aop21}.

The common approach to the simulation of viscoelastic flows couples the Navier--Stokes equations (or, in the limit of vanishing inertia, the Stokes equations) with a constitutive equation governing the evolution of the polymeric stress. The best known models of viscoelastic flows are the Oldroyd-B model and the FENE-P model, where the acronym stands for `finitely extensible nonlinear elastic with the Peterlin approximation'. Both models express the polymeric stress in terms of the conformation tensor $\mathsfbi{C}(\boldsymbol{x},t)$, defined as the tensor product of the polymer's end-to-end separation vector $\boldsymbol{R}$ by itself, averaged over all polymers contained in a volume element located at position $\boldsymbol{x}$ at time $t$, i.e., $\mathsfi{C_{ij}} \equiv \langle {R}_i {R}_j \rangle$. Clearly, the conformation tensor is defined to be positive definite. In principle, the evolution equations preserve this property, though in practice, numerical errors can lead to its loss, which in turn produces unphysical stresses and numerical instabilites. 


Simulations of elastic turbulence are particularly susceptible to numerical instabilities because the tensor field of $\mathsfbi{C}$ develops very large gradients as the polymers are stretched out by the chaotic flow (the trace $\trC$ is the squared extension of polymers). A range of special numerical schemes have been developed, therefore, to resolve these steep variations of $\mathsfbi{C}$ and to guarantee its positive definiteness \citep{aop21}. As discussed below, these methods involve a combination of non-trivial transformations or decompositions of $\mathsfbi{C}$, advanced advection schemes, and artificial diffusion. We thus have a variety of distinct methods that are capable of keeping $\mathsfbi{C}$ positive definite and producing stable numerical solutions. But do they capture the underlying physics? One would hope, given sufficient numerical resolution and converged results, to find agreement among these methods on key dynamical features of the flow. This is not the case. Rather, we show in this work that different methods yield predictions with qualitative differences in the large-scale dynamics, sufficiently significant not only to appear in averaged statistics (\textit{e.g.} kinetic energy fluctuations) but also to change the rate of scalar mixing. 

Given such substantial disagreement between methods, how is the correct one to be identified? This question is particularly difficult to answer in the context of elastic turbulence because it does not enjoy the same degree of universality as Newtonian turbulence---its statistical properties vary with the polymer concentration, the forcing, and the boundary conditions. In the absence of universal laws, any numerical solution that fluctuates chaotically without diverging appears plausible. Here, we settle this question using a mathematical result for the Oldroyd-B model which states that the determinant of the polymer conformation tensor must stay greater than unity \citep{musacchio,hl07}. This lower bound, which has not received due attention in the literature, allows us to decisively identify erroneous solutions simply by monitoring the minimum value of the determinant of $\mathsfbi{C}$. 

Before focusing on specific numerical procedures, it is helpful to survey the three classes of approaches that have been developed to stabilize simulations of the Oldroyd-B and FENE-P models. 

The first approach involves artificial diffusion. The true centre-of-mass diffusion of polymers is very weak and the corresponding diffusive length and time scales are orders of magnitude smaller than the large-scales of the flow. Thus, the computational cost of accurately resolving the diffusion of polymers (and of the associated elastic stress) is prohibitive, and so the diffusive term is omitted from the evolution equation for $\mathsfbi{C}$, rendering it hyperbolic. When elasticity is dominant (measured by large values of the non-dimensional Weissenberg number $\Wi$), the lack of a diffusive dissipation mechanism results in the formation of extremely sharp gradients in the field of $\mathsfbi{C}$, which if not treated carefully can cause a loss of positive-definiteness and numerical instability \citep[\textit{e.g.,}][]{vc03,fk05,dubief05}. A plausible remedy to this \textit{high-$\Wi$ number problem} \citep{aop21} is to re-introduce diffusion but with an artificial enhancement. 

\citet{sb95} added a Laplacian term $\kappa\Delta\mathsfbi{C}$  with a constant diffusivity $\kappa$ that was artificially increased by several orders of magnitude (whence the name `global artificial diffusivity') to stabilize viscoelastic simulations at high Reynolds numbers. The early simulations of elastic turbulence did the same \citep[\textit{e.g.},][]{bbbcm08,ts09,liu13}. 
Unfortunately, this enhanced diffusion was found to impact the  large-scale structures and dynamics of the flow, by smearing out the conformation tensor in the regions of high stretching (\citealt*{myc01}; \citealt{yk04,vrbc06,dubief05,std18}). These artifacts are particularly prominent in elastic turbulence wherein velocity fluctuations are entirely dictated by elastic forces: excessive diffusion causes polymer stress to leak into and destabilize regions of low straining which otherwise would have remained unaffected by polymer feedback \citep{gv19}. 
To attenuate these spurious effects of enhanced diffusion, one may attempt to confine its action to regions of the flow where large polymer stretching and associated numerical errors are expected. This can be achieved by switching on diffusion only at those locations where $\mathsfbi{C}$ loses positive-definiteness \citep{myc01} or by making $\kappa$ a function of either the local strain-rate \citep{gillissen19} or the gradients of $\mathsfbi{C}$ \citep*{dfs22}. Another alternative is to use `hyperdiffusivity', \textit{i.e.} a higher power of the Laplacian of $\mathsfbi{C}$, possibly in combination with a space-time dependent hyper-diffusivity coefficient \citep{gillissen19}. In spectral space, the action of localized diffusion is to selectively damp the energy of high-wavenumber modes. In fact, pseudo-spectral simulations have used spectral filters to directly suppress high-wavenumber modes and thereby curb the growth of steep gradients \citep{hl07filter,btrd11}. 

The second set of strategies preserves the positive definiteness of $\mathsfbi{C}$ by simulating suitable reformulations of the constitutive equations.
\citet{vc03} applied the Cholesky decomposition to $\mathsfbi{C}$ and evolved its Cholesky factor. Furthermore, in order to ensure the positiveness of the diagonal elements of the Cholesky factor (necessary for the uniqueness of the decomposition), they evolved equations for the logarithm of these elements. This two-step reformulation, which will henceforth be referred to as the Cholesky-log decomposition, has been used in several recent simulations (\textit{e.g.}, \citealt{gp17,pgvg17,gcmb18,gv19}; \citealt*{gcb21}; \citealt{dfs22,song22,song23}).

A logarithmic transformation, by itself, is also the basis of the log-conformation representation of \citet{fk04,fk05}, which evolves the matrix logarithm of the conformation tensor.
The polymer-stress can grow exponentially as one nears zones of high strain-rate, such as stagnation points. In such situations, polynomial approximations to $\mathsfbi{C}$ fail, but the matrix logarithm of $\mathsfbi{C}$, whose profile is simply linear, is easily resolved. Moreover, the positive definiteness of $\mathsfbi{C}$---calculated by an exponentiation---is guaranteed. The log-conformation representation is implemented in the rheoTool solver of the open-source program OpenFOAM$^\text{\textregistered}$ \citep{pa17}, which
has been used in elastic turbulence simulations of flows with boundaries, namely the cross-slot \citep*{cmb20}, the Taylor--Couette \citep*{stark18,stark20}, and the swirling von K\'arm\'an flows \citep{stark22}. 

Yet another reformulation strategy is based on the symmetric square root (SSR) decomposition  of $\mathsfbi{C}$ \citep[see also \citealt*{thomases20}]{btrd11}. This decomposition ensures positive definiteness just like the Cholesky decomposition, but has the advantage of simpler evolution equations (even when compared to the log-transformation method) which makes it easier to code. A further benefit, particularly for mathematical analysis, is that unlike $\mathsfbi{C}$ the symmetric square root of $\mathsfbi{C}$ is an element of a vector space.

The third set of approaches focuses on the numerical methods used for spatial discretization of the evolution equations, and as such complement the first two strategies. If artificial diffusion or spectral filtering is used to maintain a smooth field of $\mathsfbi{C}$, then pseudo-spectral methods are an ideal choice, especially for simple domains \citep{bbbcm08,ts09,btrd11,liu13,gcmb18,gcb21,thomases20}. If artificial diffusion is avoided, then one must use numerical techniques developed for hyperbolic equations. Typically, finite-difference or finite-volume based spatial discretizations are used \citep{vrbc06,gp17,pgvg17,gv19,aop21,lin22}, along with shock-capturing advection schemes designed to resolve the steep gradients of $\mathsfbi{C}$ (\textit{e.g.}, the advection scheme of \citet{kt00} which was adapted for polymer solutions by \citet{vrbc06}). Since viscosity keeps the velocity field smooth, some studies adopt a hybrid approach, in which a finite-difference based shock-capturing scheme is used for evolving $\mathsfbi{C}$ in combination with a lattice Boltzmann \citep{dfs22,dzanic_computers_fluids22_a,dzanic23} or a pseudo-spectral method \citep{zx20,lin22,song22,song23} for the flow. 

In this work, we compare three different reformulations of the conformation tensor, and also investigate the effect of including local polymer-stress diffusion. Calculations without artificial diffusion are facilitated by the use of the Kurganov--Tadmor advection scheme. Such schemes resolve sharp gradients up to the grid scale, and so the fine structures of the solution depend on spatial resolution; this dependence is also examined. Comparisons of different reformulations have been performed previously, for laminar flows, by \citet*{apa12}, \citet{pinho16} and \citet*{hsa21}. The current study addresses this issue in the context of elastic turbulence.
 
We begin, in \S~\ref{sect:eqs}, by presenting the Oldroyd-B and FENE-P models along with their main properties. In particular, we recall a mathematical result for the Oldroyd-B model which establishes a lower bound on the determinant of the polymer conformation tensor. While most studies in the past have focused on ensuring that the determinant of $\mathsfbi{C}$ stays greater than or equal to zero (to preserve positive definiteness), the determinant must in fact obey a more stringent bound, for the Oldroyd-B model, and remain greater than unity \citep{musacchio,hl07}. This additional constraint will be invaluable for assessing the accuracy of our numerical simulations.  

Our study is carried out in the simplified setting of a two-dimensional periodic square with cellular forcing, a configuration that develops elastic-turbulence for sufficiently high elasticity \citep{gp17,pgvg17}. Cellular forcing is particularly useful for testing and comparing simulation methods because it generates highly localized straining regions, and hence sharp gradients in the polymer extension, which can be the source of numerical inaccuracies. Another forcing that has similar features and which has also been used in simulations of elastic turbulence is the four-roll mill forcing \citep{ts09}. Further details of our simulations are provided in \S~\ref{sect:eqs}.

In \S~\ref{sect:decomposition}, we compare different reformulations of the conformation tensor, in particular the Cholesky-log and SSR decompositions. The latter is the simplest reformulation, and thus the easiest to implement in code, while the former has been widely used in recent simulations of elastic turbulence. Moreover, for the purposes of our analysis, the Cholesky-log decomposition has an advantage over the log-conformation reformulation. While the two approaches share a logarithmic transformation, it is not essential in the case of the Cholesky-log decomposition---the positive-definiteness of $\mathsfbi{C}$ is guaranteed by the Cholesky decomposition itself. So, by comparing results with and without the log transformation, we will be able to determine its influence on the accuracy of predictions. As we shall see, the conclusions of our study are sufficiently general so as to hold for the log-conformation representation as well.

Section~\ref{sect:diffusion} investigates the effect of adding \textit{local} polymer-stress diffusion to the constitutive equation. While it is now clear that \textit{global} artificial
 diffusion has a strong impact on the numerical solution and can cause spurious large-scale dynamics, the effect of localized diffusion is not yet entirely understood.
On the one hand, the intervention of local diffusion is confined to a small region of the flow. On the other hand, by acting where polymer stretching is most intense, the diffusive spreading selectively modifies the transport of the largest polymeric stresses, which exert the strongest feedback on the flow. Here we consider a form of local diffusion that has been applied recently to simulations of elastic turbulence \citep{dfs22}. 

In the absence of diffusion, the smallest length scale over which the conformation tensor may vary is determined by the spatial resolution. So, do the large-scale dynamics converge even as the gradients of $\mathsfbi{C}$ get ever sharper with increasing resolution? We address this question in \S~\ref{sect:resolution}, by examining the effects of a decreasing grid size on the flow structures and the energy spectrum.

The principal applications of elastic turbulence arise from its ability to generate chaotic mixing at low Reynolds numbers. From this perspective, it is important that numerical simulations accurately predict the mixing of a scalar. Although solving the scalar transport equation is straightforward, errors in computing the flow, especially its large-scale structures, will contaminate the predictions of mixing properties. We demonstrate this in \S~\ref{sect:mixing}, by studying the dispersal of a scalar blob, and show that different numerical treatments of the constitutive equation produce markedly different mixing behaviours.

Finally, in \S~\ref{sect:conclusions}, we summarize the findings of our study and discuss their implications for simulations of elastic turbulence.

\section{Evolution equations and numerical simulations}
\label{sect:eqs}

Consider the flow of a dilute polymer solution, whose kinematic
viscosity has contributions of $\nu$ from the solvent and $\nu_p$ from the polymer. The latter is proportional to the concentration of the polymer, which is characterized by its relaxation time to equilibrium, $\tau_p$, and the squared ratio of its contour length to the equilibrium length, \textit{i.e.} the extensibility parameter $b$. The dynamics of the solution is then described in terms of the velocity field $\boldsymbol{u}(\boldsymbol{x},t)$ and the polymer conformation tensor $\mathsfbi{C}(\boldsymbol{x},t)$, here scaled by the mean squared extension at equilibrium.
Adopting either the Oldroyd-B or the FENE-P model, and considering the limit of negligible fluid inertia, we obtain the following coupled equations for $\mathsfbi{C}(\boldsymbol{x},t)$ and $\boldsymbol{u}(\boldsymbol{x},t)$:
\begin{subequations}
\label{eq:fene-p}
\begin{gather}
\partial_t \mathsfbi{C} + \boldsymbol{u}\bcdot\bnabla\mathsfbi{C} =\mathsfbi{C}\bcdot\bnabla\boldsymbol{u} + (\bnabla\boldsymbol{u})^\top\bcdot\mathsfbi{C} -\mathsfbi{T}_p, 
\label{eq:C}
\\
\label{eq:u}
\bnabla p = \nu\Delta\boldsymbol{u} + \nu_p \bnabla\bcdot\mathsfbi{T}_p + \boldsymbol{F},
\\
\label{eq:div}
\bnabla\bcdot\boldsymbol{u}=0,
\end{gather}
\end{subequations}
where  $p$ is the ratio of pressure to the constant density of the solution, $(\bnabla\boldsymbol{u})_{ij}=\partial_i u_j$, and $\boldsymbol{F}$ is a body forcing.
The polymer contribution to the stress tensor, $\mathsfbi{T}_p$, takes the form
\begin{equation}
\mathsfbi{T}_p=\frac{1}{\tau_p}[f(r)\mathsfbi{C}-\mathsfbi{I}],
\end{equation}
where  $f(r)=1$ for the Oldroyd-B model and $f(r)=(b-d)/(b-r^2)$ for the FENE-P model.
Here $d$ is the space dimension, $r=\sqrt{\operatorname{tr}\mathsfbi{C}}$, and $\mathsfbi{I}$ is the identity matrix.

We have considered the Stokes limit for ease of computation, since we are interested in flows at low Reynolds number ($\Rey$) for which the chaotic dynamics is solely driven by elastic instabilities. The same approach was taken, for instance, in \citet{ts09}, \citet{btrd11}, and \cite{thomases20}. In any case, the difficulties encountered in numerical simulations arise from the advection of the polymer conformation tensor and would be the same if the Navier--Stokes equations (with $\Rey \lesssim 1$) were used to evolve the flow. In fact, we have repeated some of our key calculations with the Navier--Stokes equations and found that our conclusions remain unchanged (see the \href{https://math.unice.fr/~vincenzi/elastic_turb_supplement.pdf}{ supplementary material}).

\subsection{Lower bound on the determinant of $\mathsfbi{C}$}
\label{sect:lb}

The polymer conformation tensor is positive definite by construction and hence its determinant must be positive. This property is preserved by \eqref{eq:fene-p} \citep{ck12}.
However, for the Oldroyd-B model, \eqref{eq:C} has stronger implications on the determinant of~$\mathsfbi{C}$:
\cite{hl07} proved that at long times the determinant of $\mathsfbi{C}$ must satisfy the bound
\begin{equation}
\label{eq:bound-det}
\det\mathsfbi{C}\geq 1
\end{equation}
everywhere in the domain. More precisely, let us denote the value of the conformation tensor along a given Lagrangian trajectory $\boldsymbol{x_p}(t)$ as $\mathsfbi{C}(t)=\mathsfbi{C}(\boldsymbol{x_p}(t),t)$ . If there exists a time $t_\star$ such that $\det\mathsfbi{C}(t_\star)\geq 1$, then $\det\mathsfbi{C}(t)\geq 1$ for all $t > t_\star$.
If in contrast $\det\mathsfbi{C}(t_\star)< 1$, then $\det\mathsfbi{C}(t)$ keeps growing as long as $\det\mathsfbi{C}(t)< 1$.
When applied to all trajectories this result implies that, asymptotically in time, \eqref{eq:bound-det} must hold everywhere in space. The same bound was derived by \cite{musacchio} using dynamical systems theory.
Obviously, if at time $t=0$ the determinant of $\mathsfbi{C}$ is greater than or equal to unity, as is the case in most simulations and certainly in all those presented here, then it must remain so throughout the subsequent evolution. 

It is important to note that \eqref{eq:bound-det} has been proved only for the Oldroyd-B model \citep{hl07}, and a similar result that is local in space and time and uniform in $\bnabla\boldsymbol{u}$ is not available, to our knowledge, for the FENE-P model. A bound analogous to \eqref{eq:bound-det} has been derived, though, for the Giesekus model \citep{masmoudi}.

When \eqref{eq:bound-det} is combined with the inequality $\operatorname{tr}\mathsfbi{C}\geq d(\det\mathsfbi{C})^{1/d}$, which holds for any symmetric positive-definite $d\times d$ matrix, we obtain 
\begin{equation}
\label{eq:bound-tr}
\operatorname{tr}\mathsfbi{C}\geq d.
\end{equation}
This bound on the trace of $\mathsfbi{C}$ has a natural physical interpretation. Recall that $\operatorname{tr}\mathsfbi{C}$ is the squared extension of the polymer, ensemble averaged over thermal noise, and that $\operatorname{tr}\mathsfbi{C} = d$ when the polymer solution is at equilibrium (for which $\mathsfbi{C}=\mathsfbi{I}$ since $\mathsfbi{C}$ has been scaled by the mean square polymer extension at equilibrium). Hence, \eqref{eq:bound-tr} implies that an incompressible flow cannot squeeze a polymer below its mean equilibrium extension.

We shall show below that \eqref{eq:bound-det} is a useful criterion for testing the accuracy of numerical simulations.

\subsection{Matrix decompositions}

In the following sections, we will use two decompositions of the polymer conformation tensor, both of which preserve positive-definiteness.

\cite{vc03} considered the tensor $\mathsfbi{J}=f(r)\mathsfbi{C}$ and its Cholesky decomposition
$\mathsfbi{J}=\mathsfbi{L}\mathsfbi{L}^\top$, where $\mathsfbi{L}$ is a lower triangular matrix with positive diagonal entries.
The evolution equations for $\mathsfbi{L}$ in two dimensions are given in Appendix~\ref{app:matrix}. To preserve the uniqueness of the decomposition, the diagonal elements of $\mathsfbi{L}$ must remain positive during the time evolution. This is achieved by evolving $\tilde{L}_{ii}=\ln L_{ii}$  ($i=1,\dots,d$) and then exponentiating it at every time step to obtain ${L_{ii}}$. The off-diagonal elements are evolved directly. This decomposition method is termed the Cholesky-log decomposition. For reasons that become clear later (in \S~\ref{sect:decomposition}) we also consider just the Cholesky decomposition without the logarithmic transformation.

The formulation proposed by \cite{btrd11} considers the symmetric square root (SSR) decomposition $\mathsfbi{C}=\mathsfbi{B}\mathsfbi{B}^\top = \mathsfbi{B}^2$, where $\mathsfbi{B}=\mathsfbi{B}^\top$ is symmetric.
The evolution equations for $\mathsfbi{B}$ are formulated so as to preserve its symmetry and hence the uniqueness of the decomposition (see  Appendix~\ref{app:matrix} for the two-dimensional equations).

\subsection{Numerical simulations}\label{sec:nummethod}

We perform simulations in a two-dimensional ($d=2$) periodic square $V=[0,2\upi]^2$ and use a cellular forcing $\boldsymbol{F}(x,y)=f_0(-\sin Ky, \sin K x)$ to drive the flow. The wavenumber $K$ and magnitude $f_0$ of the forcing set the large length and velocity scales of the flow, $\ell = 1/K$ and  $U=f_0/\nu K^2$, which in turn yield a large turnover time scale $T=\ell/U=\nu K/f_0$.
For a Newtonian fluid ($\nu_p=0$), the velocity field $\boldsymbol{u}=-\boldsymbol{F}/\nu K^2$ is a 
solution of \eqref{eq:u}. 
If polymers are added to the flow ($\nu_p\neq 0)$ and the Weissenberg number $\textit{Wi}=\tau_p/T$ is sufficiently large, then the flow becomes chaotic and exhibits elastic turbulence 
\citep{gp17,pgvg17}. 
In our simulations we take $\nu=0.05$, $f_0=0.02$, $K=2$, and $\tau_p=50$, which yield
$\Wi=10$. In addition, $\nu_p=10^{-2}$ and $b=10^4$ (this value of $b$ has been used previously in simulations of elastic turbulence, see \textit{e.g.} \citealt{song23}).
The initial condition for the polymer conformation tensor is $\mathsfbi{C}=\mathsfbi{I}$. While we illustrate the results for this set of parameters, simulations
for other parameter values---different polymer contour lengths and forcing length scales---show that our conclusions remain unaltered \citep{sumithra}.
 
To integrate the flow equation, we use the vorticity--velocity formulation, {\it i.e.} we evolve the vorticity $\omega=(\nabla\times\boldsymbol{u})\bcdot\hat{\boldsymbol{z}}$ and calculate the velocity from the stream function $\psi=\Delta^{-1}\omega$ as $\boldsymbol{u}=(-\partial_y\psi,\partial_x\psi)$. 
Regarding the polymeric component, we solve either the equations for the symmetric square root $\mathsfbi{B}$ or those for the Cholesky factor $\mathsfbi{L}$ (see Appendix~\ref{app:matrix}).
Fourth-order central differences are used to discretize the equations for the flow and the polymer, except for the advection of the latter. For the inversion of the Laplacian operator, required to calculate $\psi$ from $\omega$, we take advantage of the periodic boundary conditions and use the Fourier pseudo-spectral method. The velocity field is then obtained from $\psi$ via finite differences. The spatial resolution is $256^2$ in all simulations, except in \S~\ref{sect:resolution}, where it is increased to $512^2$, $1024^2$, and $2048^2$. The time integration uses a second-order Runge--Kutta scheme with a time step $\delta t=2\times 10^{-3}$ for the lowest spatial resolution and $\delta t=1\times 10^{-3}$ for the higher resolutions.

The advection term in the equations for  $\mathsfbi{B}$ or $\mathsfbi{L}$ is treated according to the scheme of \cite{kt00}. This scheme requires the calculation of the velocities at the faces of grid cells, \textit{i.e.}, midway between points on a uniform grid. To preserve incompressibility, we first obtain the streamfunction at the faces, via linear interpolation from the grid points, and then use finite differences to calculate the face-velocities.
Following \cite*{pmp06} and \cite{gpp15}, we apply the Kurganov--Tadmor scheme to the advection of the factor matrix $\mathsfbi{B}$ or $\mathsfbi{L}$. Hence the slope limiting procedure that was used by \cite{vrbc06} to ensure positive definiteness is not required, and its omission yields two important benefits. First, the simulation time is significantly reduced because the eigenvalues of $\mathsfbi{C}$ need not be calculated. Second, the advection scheme remains second-order accurate throughout the computation, unlike the scheme with the slope limiter which, to maintain positive definiteness, may locally reduce the accuracy to first-order at a significant fraction of the grid points \citep{vrbc06,lin22}.

In \S~\ref{sect:diffusion} and \S~\ref{sect:mixing} we also present simulations where we implement the local polymer-stress diffusion proposed by \cite{dfs22}.
We therefore add to \eqref{eq:C} a diffusion term of the form $\kappa(\bnabla\mathsfbi{C})  \Delta\mathsfbi{C}$, where the variable diffusivity $\kappa(\bnabla\mathsfbi{C})=\bar{\kappa} Q(\boldsymbol{x},t)/Q_{\rm max}(t)$ with $Q(\boldsymbol{x},t)=\sum_{i,j=1}^2\{\sum_{q=1}^2[\nabla_q\mathsfi{C}_{ij}(\boldsymbol{x},t)]^2\}^{1/2}$ and $Q_{\rm max}=\max_{\boldsymbol{x}\in V} Q(\boldsymbol{x},t)$.
Thus, the artificial diffusivity varies locally between zero and $\bar{\kappa}$ according to the magnitude of the derivatives of the conformation tensor. 
Following \cite{dfs22}, we take $\bar{\kappa}=5\times 10^{-5}$, which yields a P\'eclet number $\Pen=U \ell/\bar{\kappa}=10^3$. (As discussed further in \S~\ref{sect:diffusion}, this value of $\Pen$, which is the ratio of convective to diffusive transport, is two to seven orders of magnitude smaller than that encountered in experiments of elastic turbulence.) 
In these simulations, we use the Cholesky-log reformulation of the constitutive equation;  the evolution equation for the Cholesky factor $\mathsfbi{L}$ in the presence of stress diffusion can be found in \cite{dzanic_computers_fluids22_a}.

\section{Matrix decompositions and erroneous large-scale dynamics}
\label{sect:decomposition}

We begin by comparing the results obtained using different decompositions of $\mathsfbi{C}$. The simulations in this section are devoid of polymer-stress diffusion, \textit{i.e.} $\Pen=\infty$, and use the Oldroyd-B model which makes the bound \eqref{eq:bound-det} available for testing the accuracy of the results.

\begin{figure}
\centering
\begin{subfigure}[b]{\textwidth/2}
\centering
\includegraphics[width=.85\textwidth]{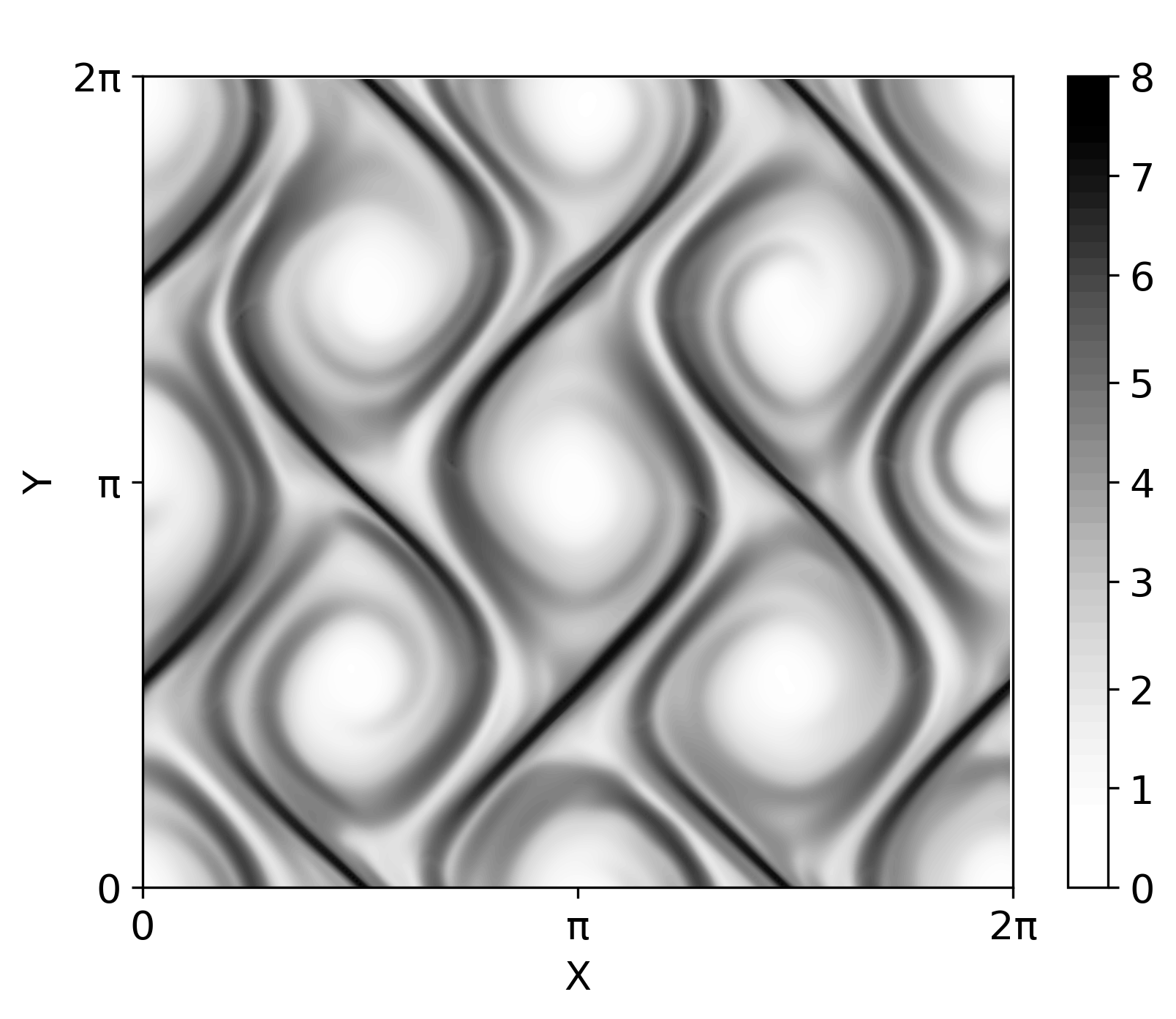}%
\label{OK}
\caption{}
\end{subfigure}%
\hfill
\begin{subfigure}[b]{\textwidth/2}
\centering
\includegraphics[width=.85\textwidth]{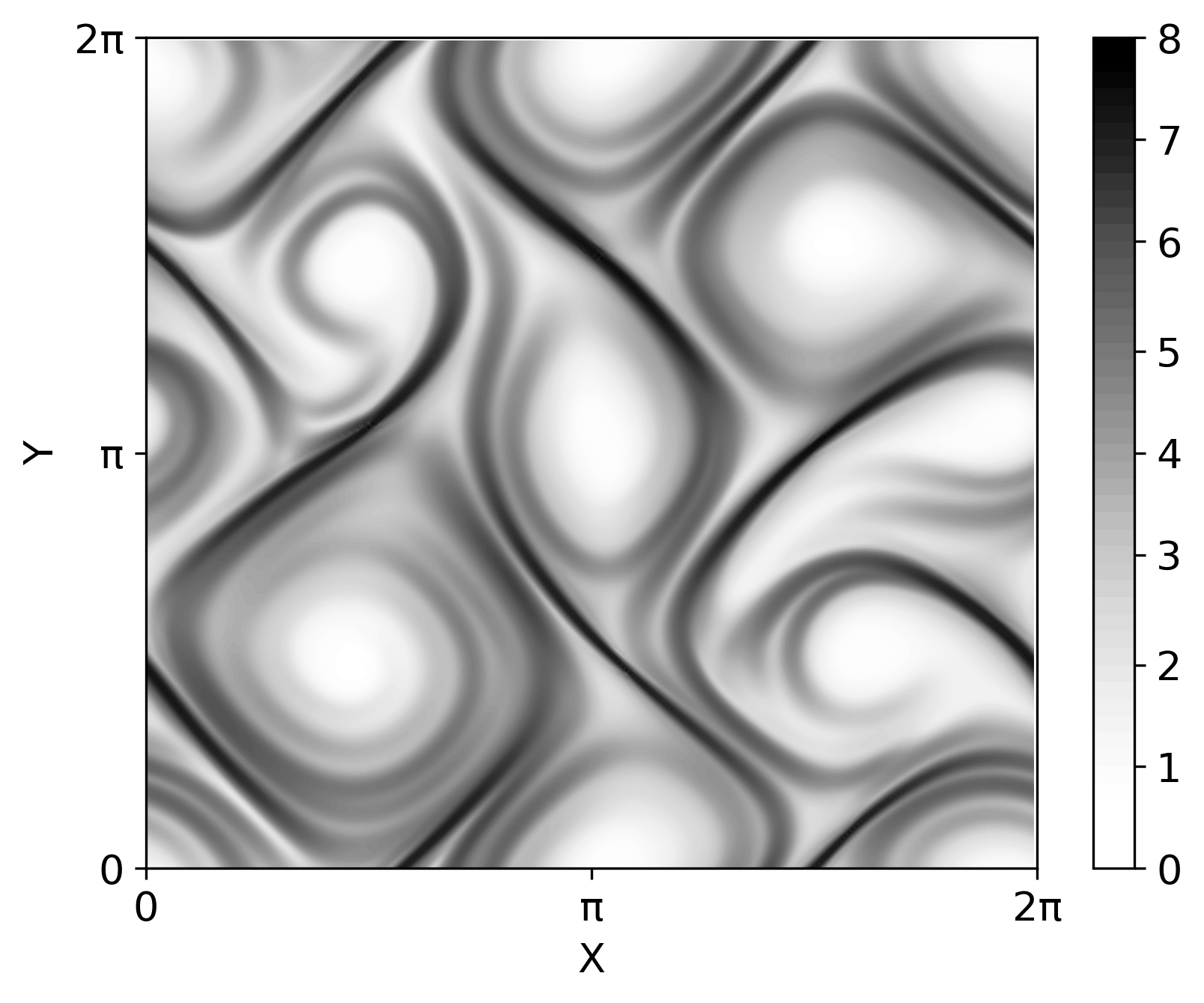}%
\caption{}
\end{subfigure}%
\caption{Representative snapshots (instantaneous) of the logarithm of $\trC$ for (\textit{a}) the Cholesky-log, and (\textit{b})~the SSR decompositions. For the corresponding animations, see supplementary \href{https://math.unice.fr/~vincenzi/Movie1.mp4}{movie 1}. Both simulations use the Oldroyd-B model.}
\label{fig:snapshots}
\end{figure}

Figures~\ref{fig:snapshots}(\textit{a}) and \ref{fig:snapshots}(\textit{b}) compare two representative snapshots of $\operatorname{tr}\mathsfbi{C}$ (the squared extension of the polymer) for the Cholesky-log and the SSR decompositions (the corresponding animations are available in supplementary \href{https://math.unice.fr/~vincenzi/Movie1.mp4}{movie 1}). 
The cellular forcing produces several large vortical cells, wherein the polymer is coiled and $\operatorname{tr}\mathsfbi{C}$ is small, separated by straining zones which give rise to thin filamentary regions, wherein the polymer is strongly stretched and $\operatorname{tr}\mathsfbi{C}$ is large. 
While the vortical cells are perturbed by the chaotic flow, in both simulations, it is clear that the symmetry of the forcing structure is much more closely preserved in the case of the Cholesky-log simulation, as compared to the SSR simulation. The former exhibits a well-ordered lattice of vortical cells, all of which maintain nearly the same orientation and shape throughout the simulation (figure~\ref{fig:snapshots}\textit{a}). In contrast, the vortical cells of the SSR simulation constantly change their orientation, shape, and size, as from time to time some cells expand while others shrink (figure~\ref{fig:snapshots}\textit{b}).
%

\begin{figure}
\centering
\begin{subfigure}[b]{\textwidth/2}
\centering
\includegraphics[width=.85\textwidth]{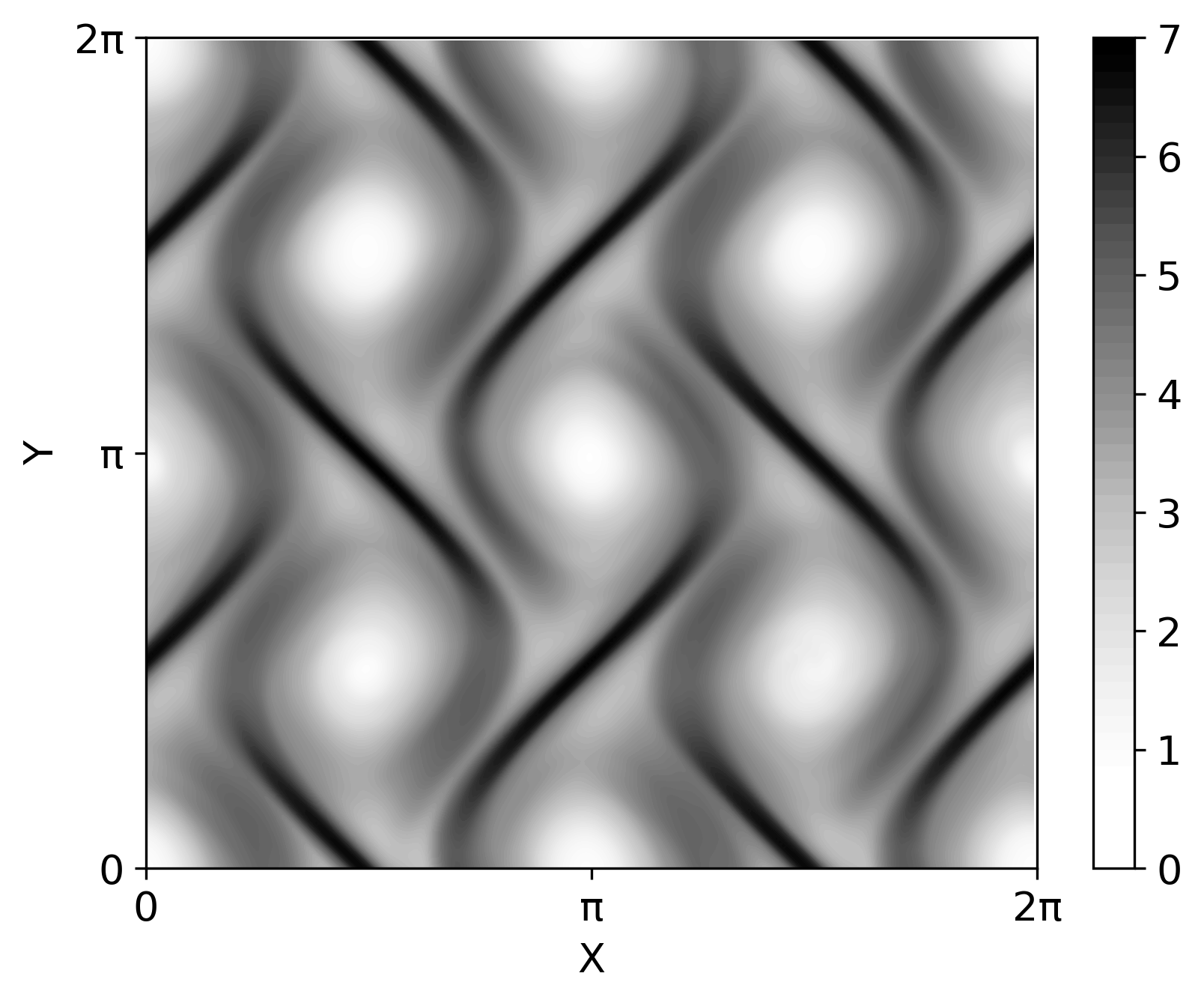}%
\caption{}
\end{subfigure}%
\hfill
\begin{subfigure}[b]{\textwidth/2}
\centering
\includegraphics[width=.85\textwidth]{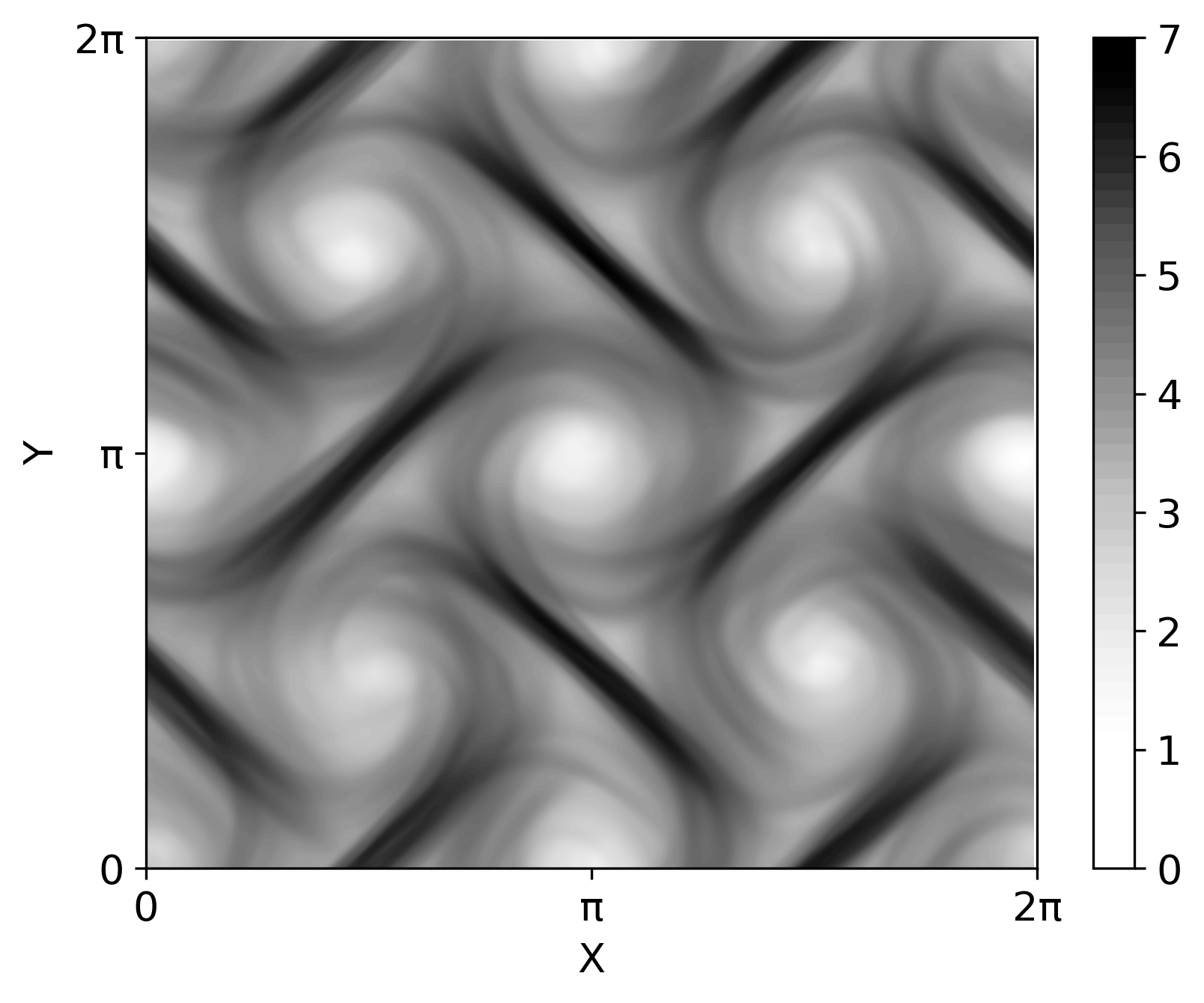}%
\caption{}
\end{subfigure}%
\caption{Time-averaged fields of the logarithm of $\trC$ for (\textit{a}) the Cholesky-log and (\textit{b})~the SSR decompositions. Both simulations use the Oldroyd-B model.}
\label{fig:mean_tr}
\end{figure}

These differences in the large-scale structures of the field of $\trC$ are not merely momentary but persist over time, as evidenced by the time-averaged fields presented in figure~\ref{fig:mean_tr}. The plots for the two decompositions are strikingly dissimilar. The time-averaged field for the Cholesky-log simulation closely resembles the corresponding instantaneous field (compare figures \ref{fig:mean_tr}\textit{a} and \ref{fig:snapshots}\textit{a}). The straining zones experience small chaotic oscillations and so are slightly smeared out in the time-averaged plot. In the SSR simulation, the time-averaged field looks very different from its instantaneous snapshot (figures \ref{fig:mean_tr}\textit{b} and \ref{fig:snapshots}\textit{b}); the jostling of the vortical cells produces a time-averaged picture of smeared cells with a high degree of symmetry unseen at any instant of time.  

\begin{figure}
\centering
\begin{subfigure}[b]{\textwidth/2}
\centering
\includegraphics[width=.8\textwidth]{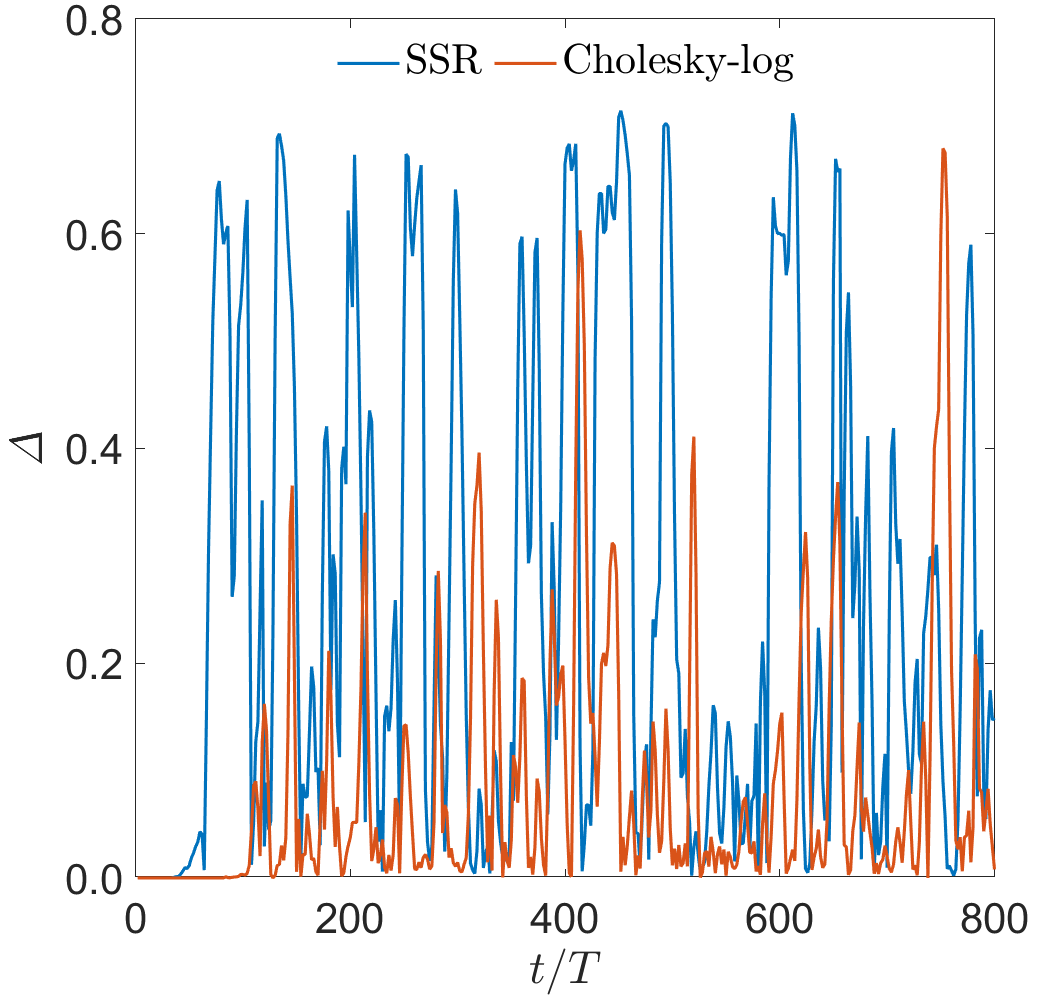}%
\caption{}
\end{subfigure}%
\hfill
\begin{subfigure}[b]{\textwidth/2}
\centering
\includegraphics[width=.8\textwidth]{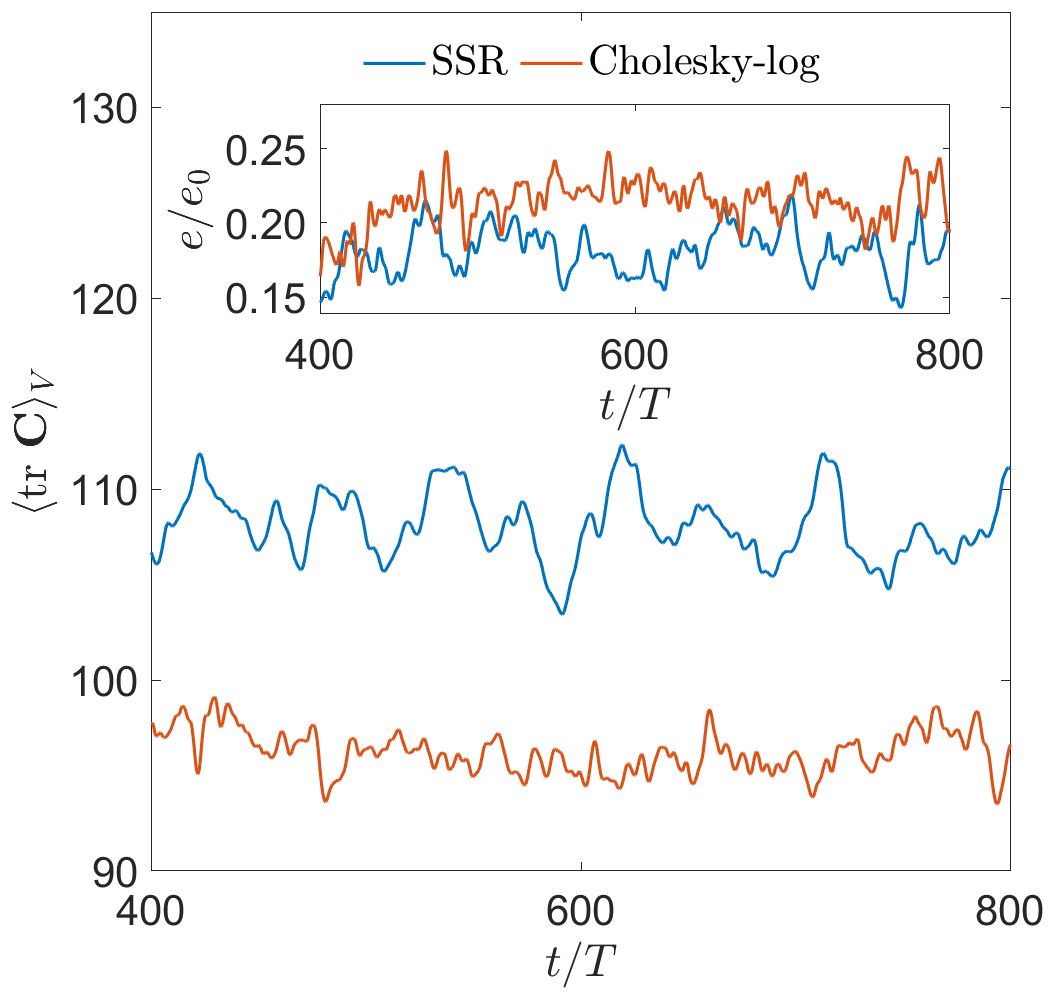}%
\caption{}
\end{subfigure}%
\caption{(\textit{a}) Time series of  $\varDelta$ (defined in \eqref{eq:Delta}); (\textit{b}) time series of the space averages of $\trC$ (main panel) and the kinetic energy $e(t)$ scaled by its Newtonian value $e_0=\frac{1}{2}U^2$ (inset). All plots refer to simulations of the Oldroyd-B model.}
\label{fig:timeseries}
\end{figure}

As a simple measure of the extent to which the cellular structure is perturbed during the evolution, we introduce the following diagnostic:
\begin{equation}
\varDelta(t)=\left\lvert\dfrac{\ln[\trC((0,0),t)]-\ln[\trC((\pi,0),t)]}{\ln[\trC((0,0),t)]+\ln[\trC((\pi,0),t)]}\right\rvert.
\label{eq:Delta}
\end{equation}
This quantity will be zero if the $\trC$ field perfectly preserves the cellular structure of the periodic forcing, which has a wavenumber $K = 2$. Therefore, $\varDelta$ is indicative of the extent of distortion of the cellular structure, albeit at a single point. The time series of $\varDelta(t)$ is plotted in figure~\ref{fig:timeseries}(\textit{a}). Clearly, the fluctuations of $\varDelta(t)$ are significantly stronger for the SSR decomposition than for the Cholesky-log decomposition.


The distinct large-scale dynamics of the two simulations lead to differences in space-averaged statistics as well. Figure~\ref{fig:timeseries}\textit{b} compares the mean polymer stretching, given by the spatial average of the trace of the conformation tensor $\langle \trC \rangle_{V} = (2\upi)^{-2}\int_V \trC \, \mathrm{d}\boldsymbol{x}$, and shows that it is higher for the SSR simulation. Higher stretching in the straining zones endows the solution with a higher extensional viscosity \citep{larsonbook}, which in turn results in the SSR simulation having a lower mean kinetic energy $e(t)=\frac{1}{2}\langle \boldsymbol{u}\cdot\boldsymbol{u}\rangle_{V}$, as shown in the inset of figure~\ref{fig:timeseries}\textit{b}.

It is important to note that the differences in the predictions of the two decompositions are independent of resolution. We have found that increasing the resolution from $256^2$ to $1024^2$ produces sharper stretching zones in both simulations but no qualitative change in the large-scale dynamics; all the differences discussed above persist.

\begin{figure}
\centering
\begin{subfigure}[b]{\textwidth/2}
\centering
\includegraphics[width=.8\textwidth]{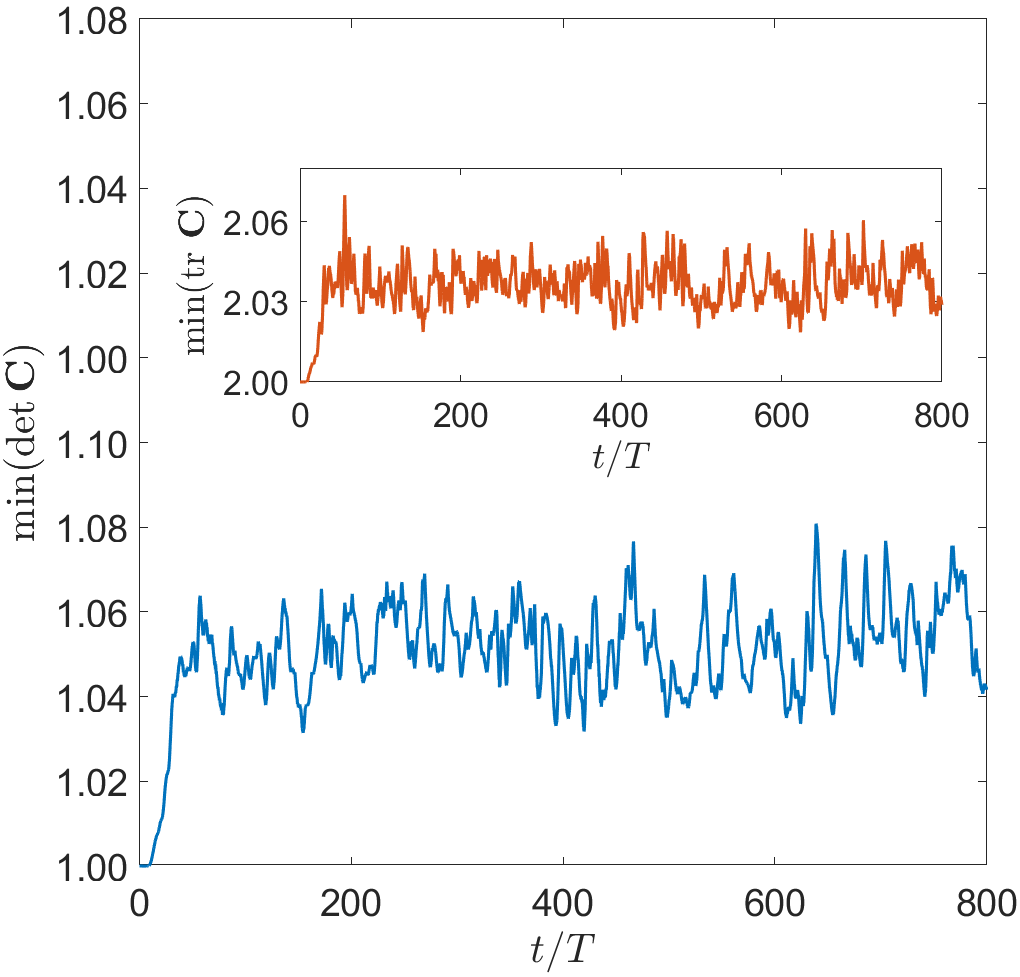}%
\caption{}
\end{subfigure}%
\hfill%
\begin{subfigure}[b]{\textwidth/2}
\centering
\includegraphics[width=.8\textwidth]{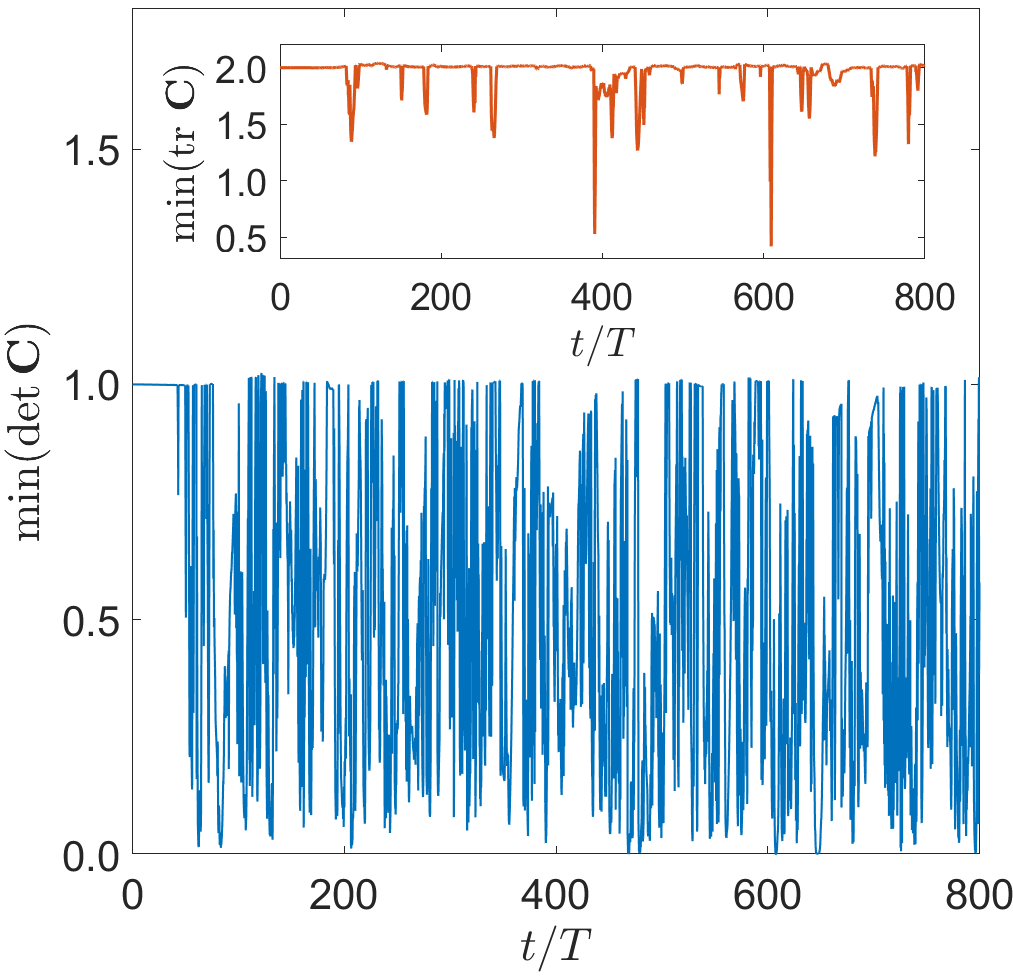}%
\caption{}
\end{subfigure}%
\caption{Time series of the minimum of $\detC$ over the domain for (\textit{a}) the Cholesky-log, and (\textit{b})~the SSR decompositions. The bound $\detC \geq 1$ (see \eqref{eq:bound-det}) is satisfied by the former simulation but is repeatedly and strongly violated by the latter. The insets show the corresponding time series of $\trC$, which as a consequence of \eqref{eq:bound-det} must remain greater than two. All plots refer to simulations of the Oldroyd-B model.}
\label{fig:det}
\end{figure}

So which of the two simulations is correct? This question is difficult to address in the present context because, as discussed in the introduction, elastic turbulence depends on the specific setting and hence lacks universal laws against which numerical predictions may be tested. In case of the Oldroyd-B model, though, the lower bound $\detC \ge 1$ presented in \S~\ref{sect:lb} comes to our aid. The time series of the minimum of $\detC$ over the domain is presented in figure~\ref{fig:det}. We see that the lower bound is respected throughout the Cholesky-log simulation (figure~\ref{fig:det}\textit{a}), whereas it is frequently violated in the SSR simulation (figure~\ref{fig:det}\textit{b}). While such violations are found to occur over a small fraction of the domain (approximately 0.1\%), they are very frequent in time and very strong---we see many instances where $\detC$ approaches zero. This leads us to conclude that the SSR simulation is erroneous and its distinguishing features, including the distortion of the cellular structure, is a consequence of inaccuracies in evolving the polymer stresses. 

In \S~\ref{sect:lb}, we recalled that the lower bound on $\detC$ implies a lower bound on the squared extension, $\trC \ge 2$. Thus, a simulation that respects the bound on $\detC$ must necessarily respect the bound on $\trC$ as well. This is seen in the inset of figure~\ref{fig:det}(\textit{a}), which presents the time series of the minimum of $\trC$ for the Cholesky-log simulation. In contrast, the SSR simulation violates both bounds (figure~\ref{fig:det}\textit{b}), though the instances of $\trC$ falling below 2 are much fewer than that of $\detC$ falling below 1. So, while $\trC$ has a simple physical interpretation and is easier to compute than $\detC$, we must verify the bound on the latter when ascertaining the accuracy of a simulation.

\begin{figure}
\centering
\begin{subfigure}[b]{\textwidth/3}
\centering
\includegraphics[width=.95\textwidth]{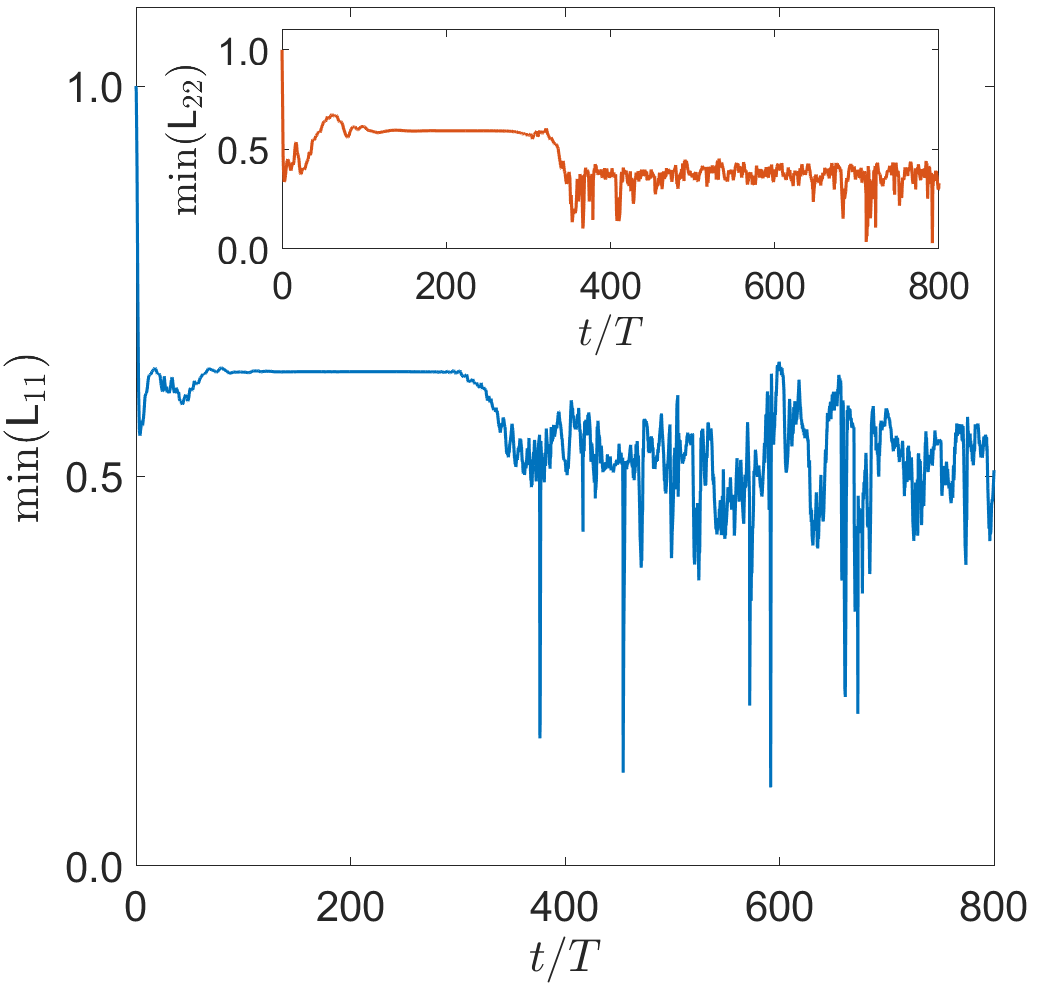}%
\caption{}
\end{subfigure}%
\begin{subfigure}[b]{\textwidth/3}
\centering
\includegraphics[width=\textwidth]{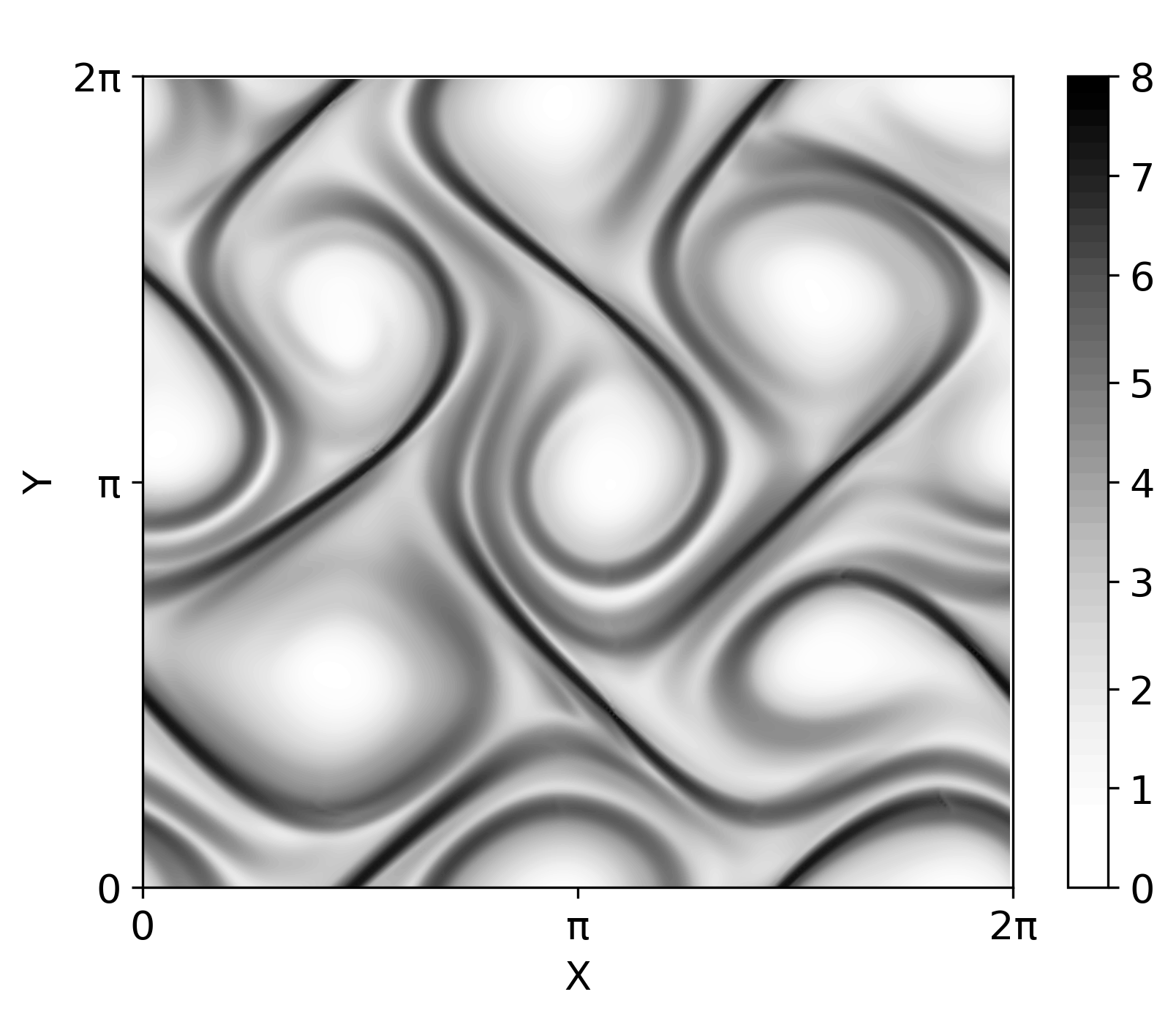}%
\caption{}
\end{subfigure}%
\begin{subfigure}[b]{\textwidth/3}
\centering
\includegraphics[width=\textwidth]{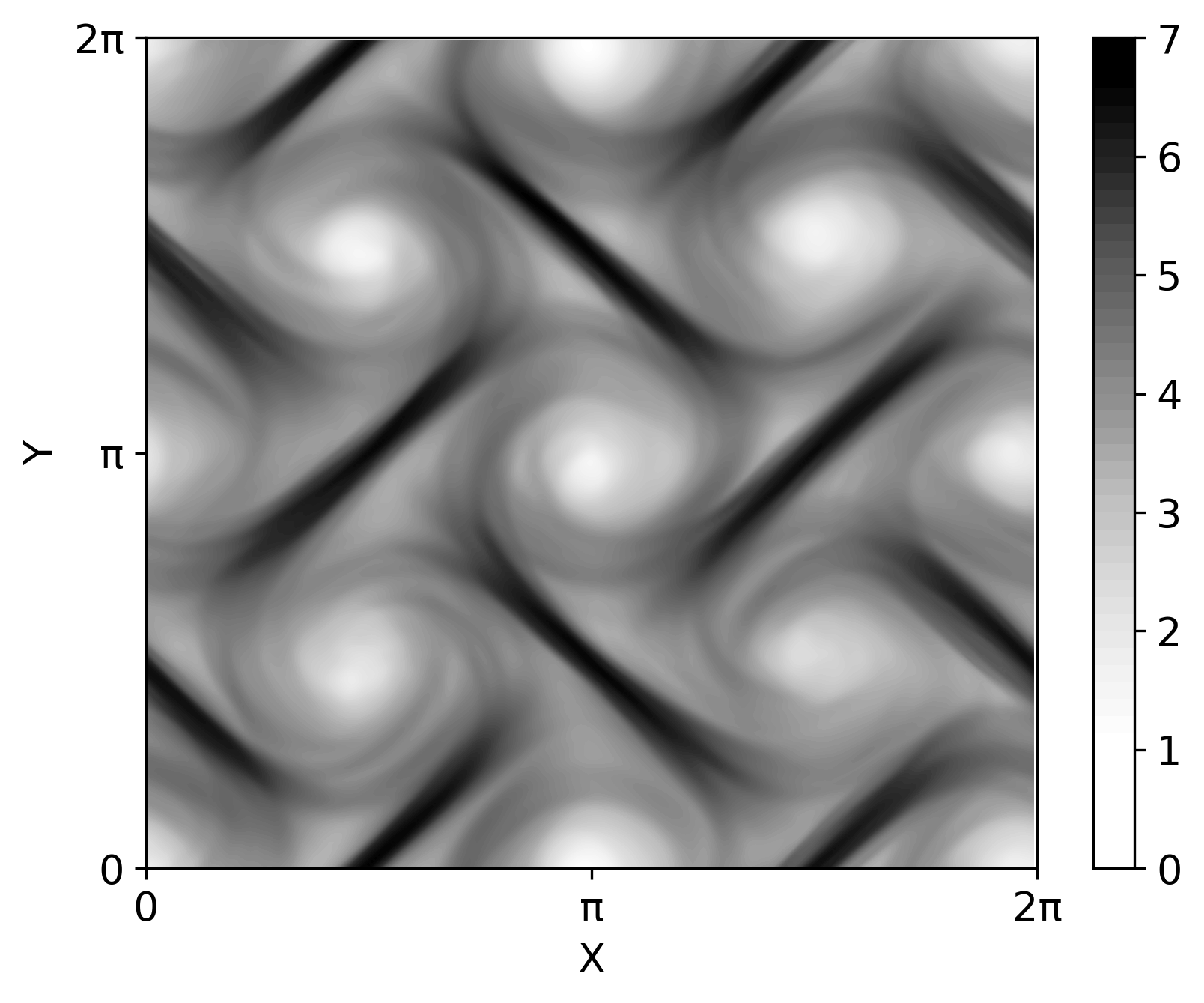}%
\caption{}
\end{subfigure}%

\bigskip
\begin{subfigure}[b]{\textwidth/3}
\centering
\includegraphics[width=.95\textwidth]{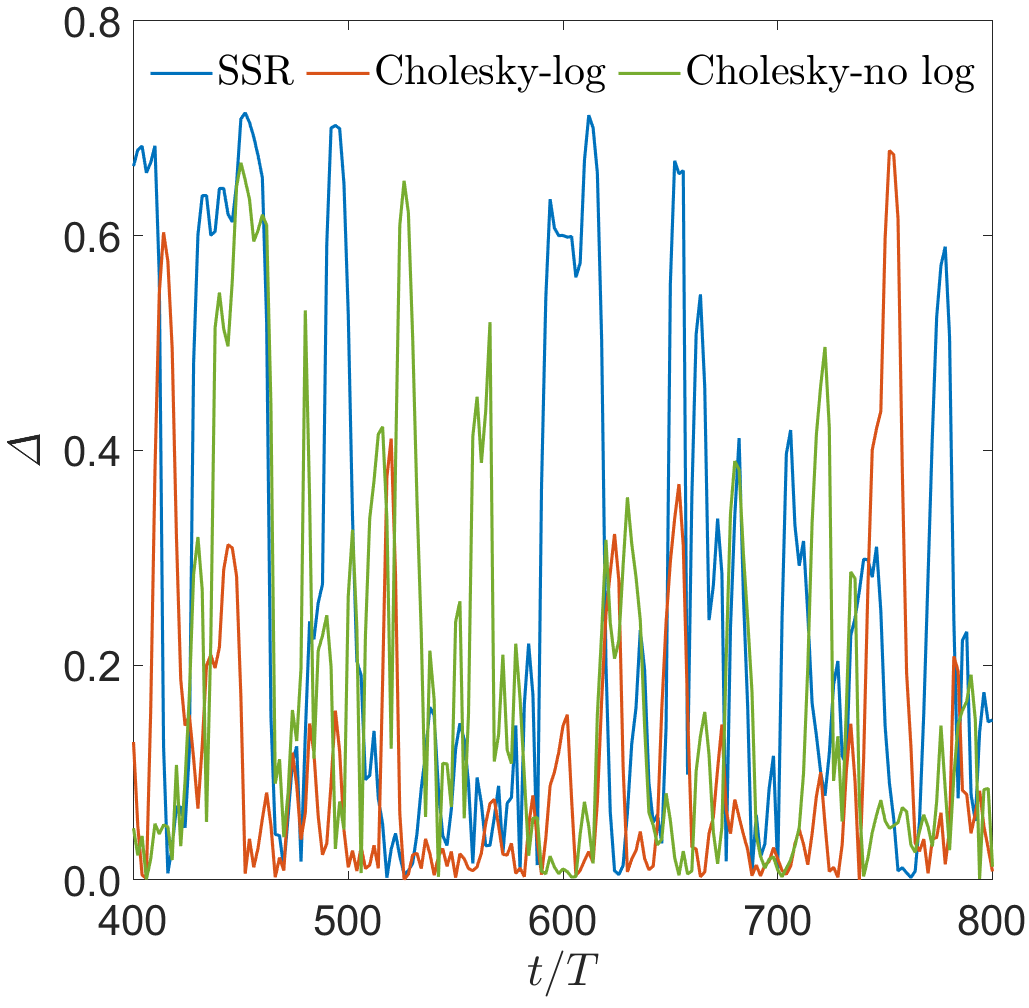}%
\label{OK}
\caption{}
\end{subfigure}%
\begin{subfigure}[b]{\textwidth/3}
\centering
\includegraphics[width=.95\textwidth]{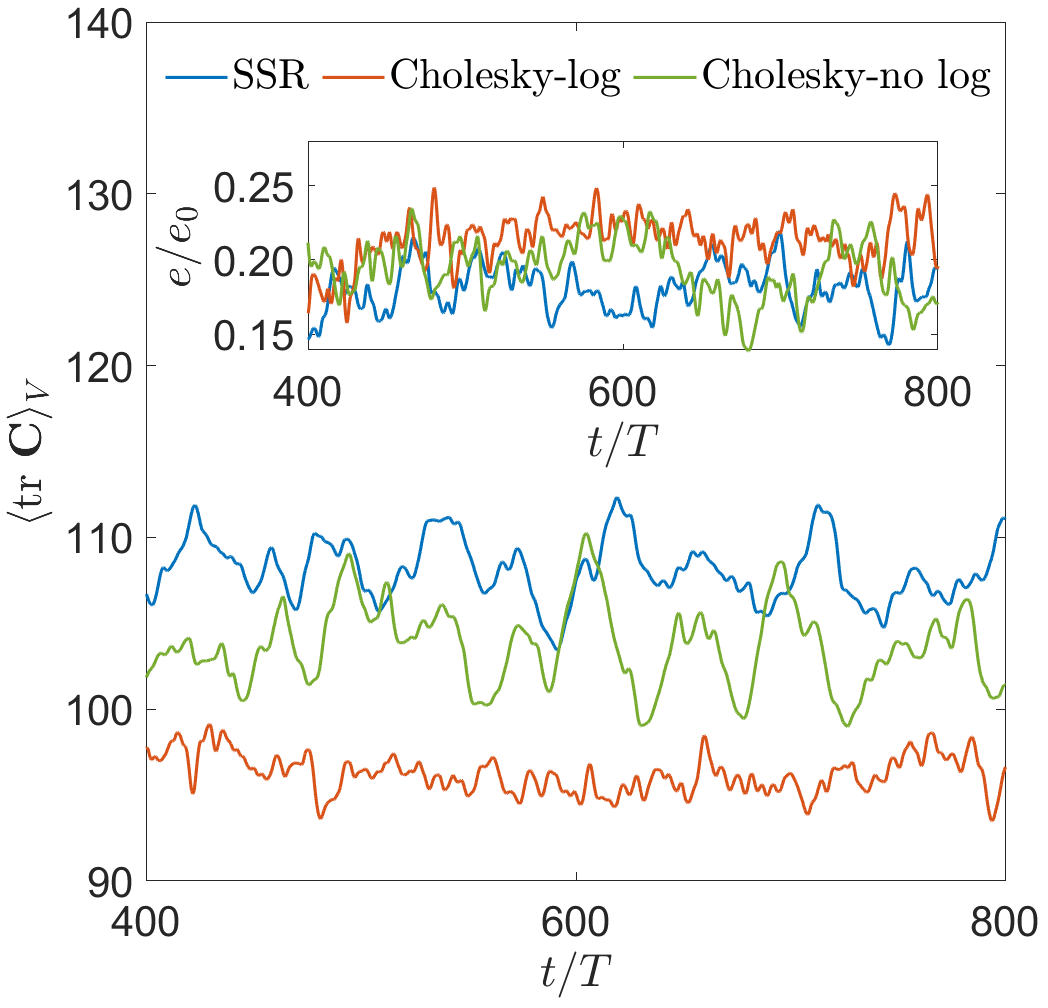}%
\caption{}
\end{subfigure}%
\begin{subfigure}[b]{\textwidth/3}
\centering
\includegraphics[width=.95\textwidth]{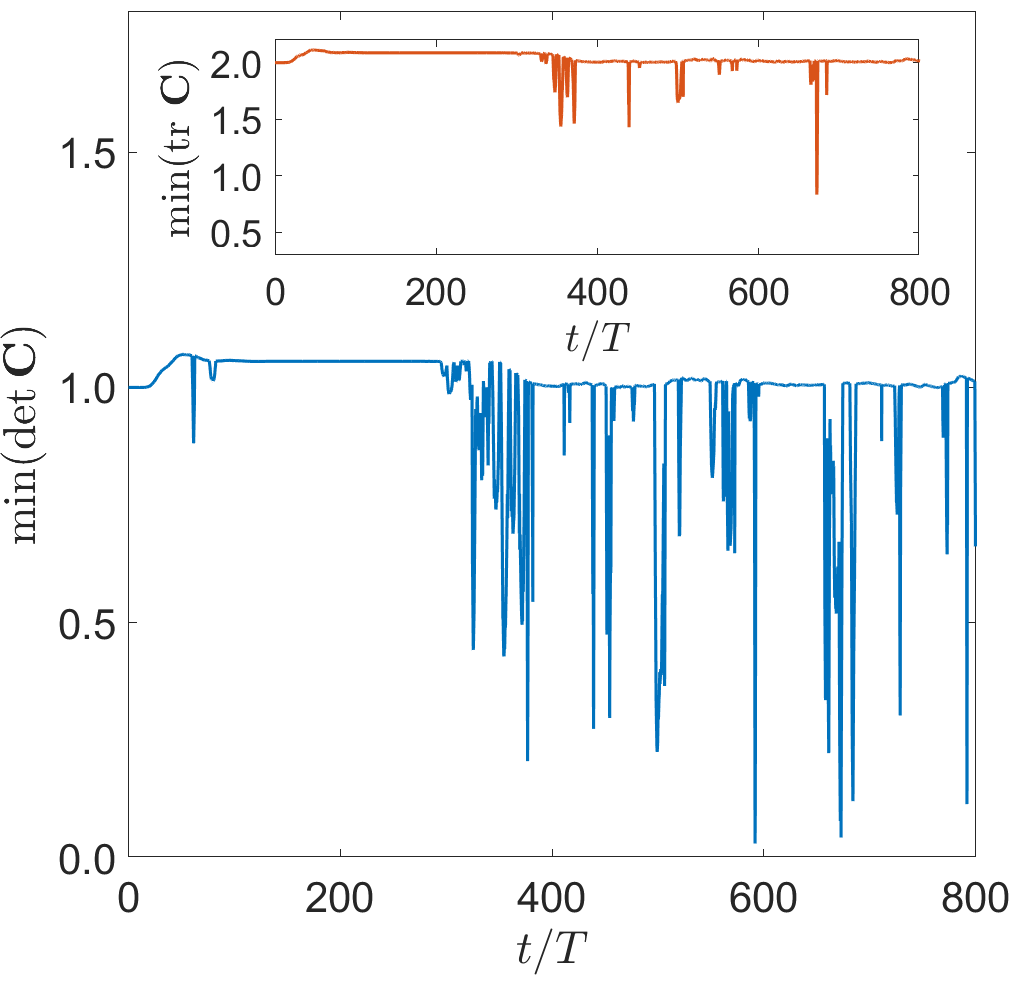}%
\label{OK}
\caption{}
\end{subfigure}%
\caption{Results for the Cholesky decomposition without the logarithmic transformation. (\textit{a}) Time series of the minimum of $\mathsfi{L}_{11}$  and $\mathsfi{L}_{22}$ (inset) over the domain;  (\textit{b}) representative snapshot (instantaneous) of the logarithm of $\trC$ (the corresponding animation is in supplementary \href{https://math.unice.fr/~vincenzi/Movie2.mp4}{movie 2}); (\textit{c}) time averaged field of the logarithm of $\trC$; (\textit{d}) time series of $\varDelta$; (\textit{e}) time series of the space averages of $\trC$ (main panel) and the kinetic energy scaled by its Newtonian value $e(t)/e_0$ (inset); (\textit{f}) time series of the minimum of $\detC$ over the domain, with the corresponding time series of $\trC$ in the inset. In panels (\textit{d}) and (\textit{e}), data for the Cholesky-log and SSR decompositions are reproduced from figure~\ref{fig:timeseries} for ease of comparison. All plots refer to simulations of the Oldroyd-B model.}
\label{fig:choleskynolog}
\end{figure}

The next question is why does the SSR simulation suffer from inaccuracies and produce predictions which are different from the accurate Cholesky-log simulation. While the two matrix decompositions are similar in spirit, they do differ on one important point---the use of a logarithmic transformation. Recall that evolving $\ln\mathsfi{L}_{ii}$ and recovering $\mathsfi{L}_{ii}$ through exponentiation is a strategy to enforce the positivity of the diagonal elements of $\mathsfbi{L}$, in order to ensure the uniqueness of the Cholesky decomposition. From a mathematical point of view, however, this step is not needed if the diagonal elements stay positive under direct evolution. To assess the role of the logarithmic transformation, we have thus performed simulations of the Cholesky decomposition without the logarithmic transformation, \textit{i.e.} we have evolved \eqref{eq:cholesky} directly. Figure~\ref{fig:choleskynolog}(\textit{a}) shows that the diagonal elements of $\mathsfbi{L}$ stay positive, nevertheless, and hence the decomposition remains unique throughout the simulation. So, we can now compare the results of the Cholesky decomposition, with and without the logarithmic transformation.

Remarkably, we find that when the logarithmic transformation is not used the results of the Cholesky simulation are analogous to those obtained with the SSR decomposition. This is apparent for all the observables that we have examined in this section:
the instantaneous and time-averaged snapshots of the logarithm of $\operatorname{tr}\mathsfbi{C}$ (figures~\ref{fig:choleskynolog}\textit{b} and~\ref{fig:choleskynolog}\textit{c}, respectively) as well as the time series of $\varDelta$, $\trC$, and the kinetic energy (figures~\ref{fig:choleskynolog}\textit{d}-\textit{e}). Moreover, if the logarithmic transformation is removed, $\detC$ frequently drops well below unity in the Cholesky simulations as well (figures~\ref{fig:choleskynolog}\textit{f}). Clearly, the higher accuracy of the Cholesky-log decomposition compared to the SSR decomposition is due to the fact that the former evolves the logarithm of the diagonal elements of the Cholesky factor.
This analysis, therefore, supports the use of a logarithmic transformation of the polymer conformation tensor for high-$\Wi$ simulations of elastic turbulence, which are characterized by large polymer-stress gradients. This conclusion is consistent with a previous comparative study on laminar flows, by \citet{apa12}, which found that simulations using the log-conformation approach remained stable up to higher values of $\Wi$ compared to the SSR decomposition.

Strictly speaking, the diagnosis carried out in this section applies only to the Oldroyd-B model, because it is based on the bound \eqref{eq:bound-det}. Nonetheless, the reason for the greater accuracy of the Cholesky-log decomposition, namely its superior ability to resolve large gradients in the fields of polymer extension and stress, is sufficiently general to expect our conclusions to hold for other models including FENE-P. Indeed, numerical simulations of the FENE-P model with the SSR and the Cholesky-log decompositions exhibit the same differences as the simulations of the Oldroyd-B model (see Appendix~\ref{app:fenep}).

\section{Artifacts of local diffusion of polymer stress}
\label{sect:diffusion}

The centres-of-mass of dissolved polymers undergo Brownian motion which produces a diffusion of polymeric stress \citep{kl1989}. However, this diffusion is much weaker than advection and, hence, becomes significant only at extremely small scales---much below what can be resolved in a practical simulation. For a flow characterized by a large-scale velocity $U$ and length $\ell$, 
the computational grid size $\delta_x$ must be small enough for diffusion to balance advection at the grid-scale: $\kappa/\delta_x^2 \sim U/\delta_x$.
Thus, the minimum required number of grid points, along any dimension, scales with the P\'eclet number, that is $\ell/\delta_x \sim U \ell/\kappa = \Pen$. 
Note that the P\'eclet number is the product of the Reynolds ($\Rey = U \ell/\nu$) and Schmidt ($\Sc = \nu/\kappa$) numbers: $\Pen = \Rey \Sc$. 

\begin{table}
  \begin{center}
\phantom{~}\noindent
  \begin{tabular}{p{0.35\textwidth}>{\centering}p{0.1\textwidth}>{\centering}p{0.08\textwidth}>{\centering}p{0.08\textwidth}>{\centering\arraybackslash}p{0.08\textwidth}}
      Reference  & $U$ (\si{m.s^{-1}}) & $\Rey$   & $\Sc$    & $\Pen$     \\[3pt]
       \citet{vs2019}   & $10^{-5}$         & $10^{-4}$     & $10^9$    & $10^5$    \\
       \citet{js2021}   & $10^{-3}$         & $10^{-2}$     & $10^9$    & $10^7$    \\
       \citet{gs2004}   & $10^{-2}$         & $1$           & $10^{10}$ & $10^{10}$    \\
  \end{tabular}
  \caption{Order of magnitude of the P\'eclet number and other key parameters in some experiments of elastic turbulence. These experiments share a geometric length scale, $\ell \sim$ \qty{e-3}{m}, and use a similar water-sucrose solvent with a density $\rho \sim$ \qty{e3}{kg.m^{-3}} and a dynamic viscosity $\mu \sim$ \qty{e-1}{Pa.s} (a hundred times more viscous than water). The third row refers to the rotating plate geometry in \citet{gs2004}. }
  \label{tab:Steinberg}
  \end{center}
\end{table}

Using order of magnitude estimates for a high Reynolds number flow ($\Rey \sim 10^3$) of a dilute aqueous polymer solution (dynamic viscosity $\mu \sim 10^{-3}$, density $\rho \sim 10{^3}$, and diffusivity $\kappa \sim 10^{-12}$, all in S.I. units), one obtains $\Sc \sim 10^6$ and hence $\Pen \sim 10^9$ \citep{kl1989}. This is much too large for a realistic simulation; indeed, numerical studies of high-$\Rey$ viscoelastic turbulence that have included diffusion typically take $\Sc \sim 1$ and thus $\Pen \sim 10^3$ \citep{sbh97,sb95,dubief05} or at most $\Sc \sim 10^2$ \citep{std18}.

Now consider elastic turbulence, which has been experimentally observed in small channels and with high-viscosity solvents. While these conditions suppress inertia, as intended, and yield small values of the Reynolds number, $10^{-4} \lesssim \Rey \lesssim 1$, they also produce very large values of the Schmidt number, $10^9 \lesssim \Sc \lesssim 10^{10}$ (recall that the diffusivity varies inversely with the solvent viscosity). Thus, once again, we obtain very large values of the P\'eclet number, $10^{5} \lesssim \Pen \lesssim 10^{10}$. Three representative examples, from the extensive experiments of Steinberg and coworkers, are given in Table~\ref{tab:Steinberg}. 
The upper range of these $\Pen$ values are too large for simulations, though the lowest value of $\Pen = 10^5$ has recently been attained \citep{morozov22}. Most simulations of elastic turbulence that include polymer stress diffusion have, however, been limited to $\Pen$ not exceeding $10^3$, thereby enhancing diffusion by several orders of magnitude \citep{ts09,liu13}.
Recently, a linear stability analysis by \citet*{bpk23} has considered $\Pen$ as large as $10^6$ and shown that polymer-stress diffusion induces a novel instability, which appears to persist in the limit of $\Pen \to \infty$. Their nonlinear simulations, which show the transition to elastic turbulence, were limited to $\Pen = 10^3$. 


\begin{figure}
\centering
\begin{subfigure}[b]{\textwidth/2}
\centering
\includegraphics[width=.85\textwidth]{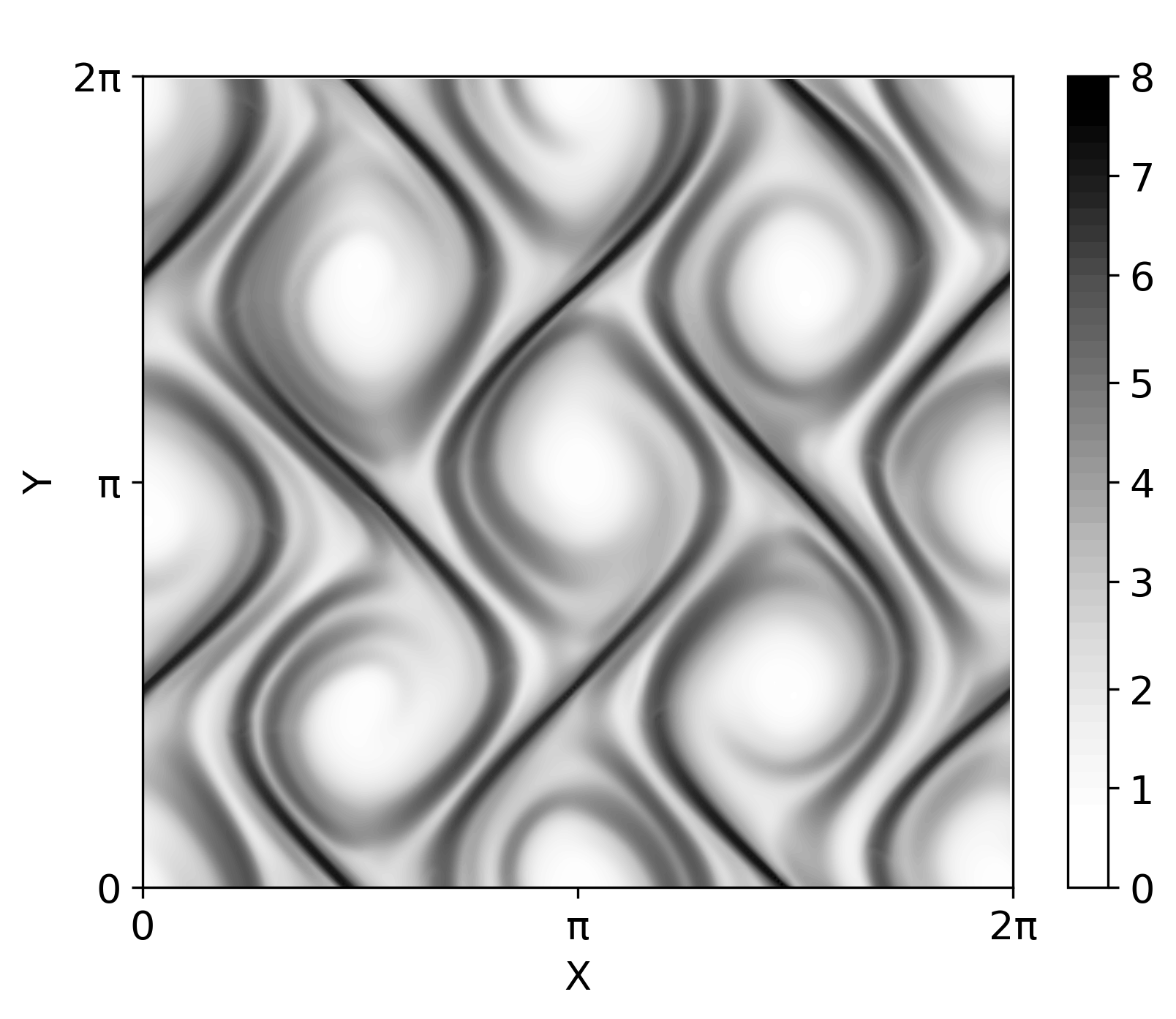}%
\caption{}
\end{subfigure}%
\hfill
\begin{subfigure}[b]{\textwidth/2}
\centering
\includegraphics[width=.85\textwidth]{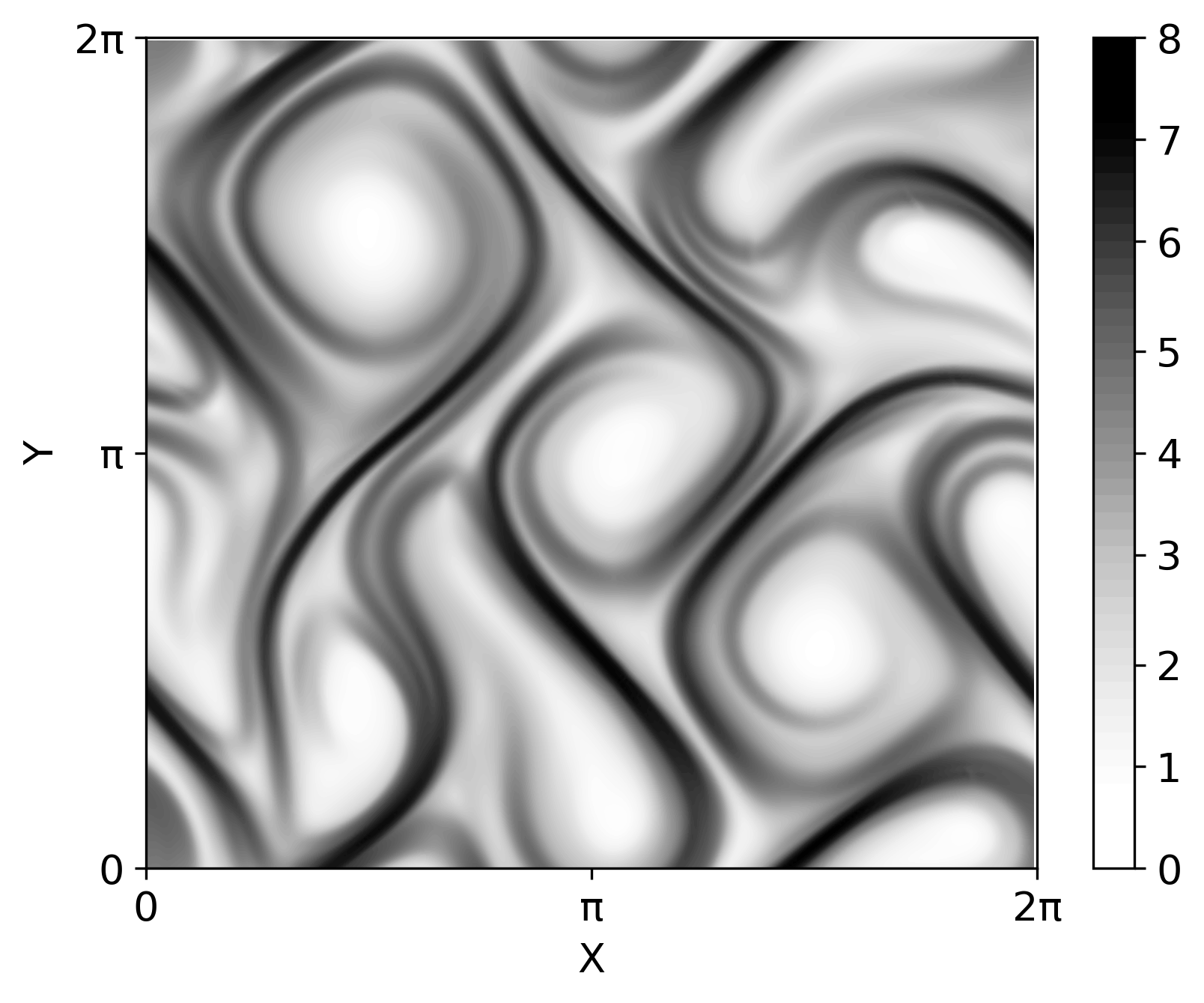}%
\caption{}
\end{subfigure}%
\caption{Representative snapshots of the logarithm of $\trC$ for the Cholesky-log reformulation of the FENE-P model with (\textit{a}) $\Pen=\infty$ and (\textit{b}) $\Pen=10^3$. For the corresponding animations, see supplementary \href{https://math.unice.fr/~vincenzi/Movie3.mp4}{movie 3}.}
\label{fig:snapshots-diffusion}
\end{figure}
\begin{figure}
\centering
\begin{subfigure}[b]{\textwidth/3}
\includegraphics[width=\textwidth]{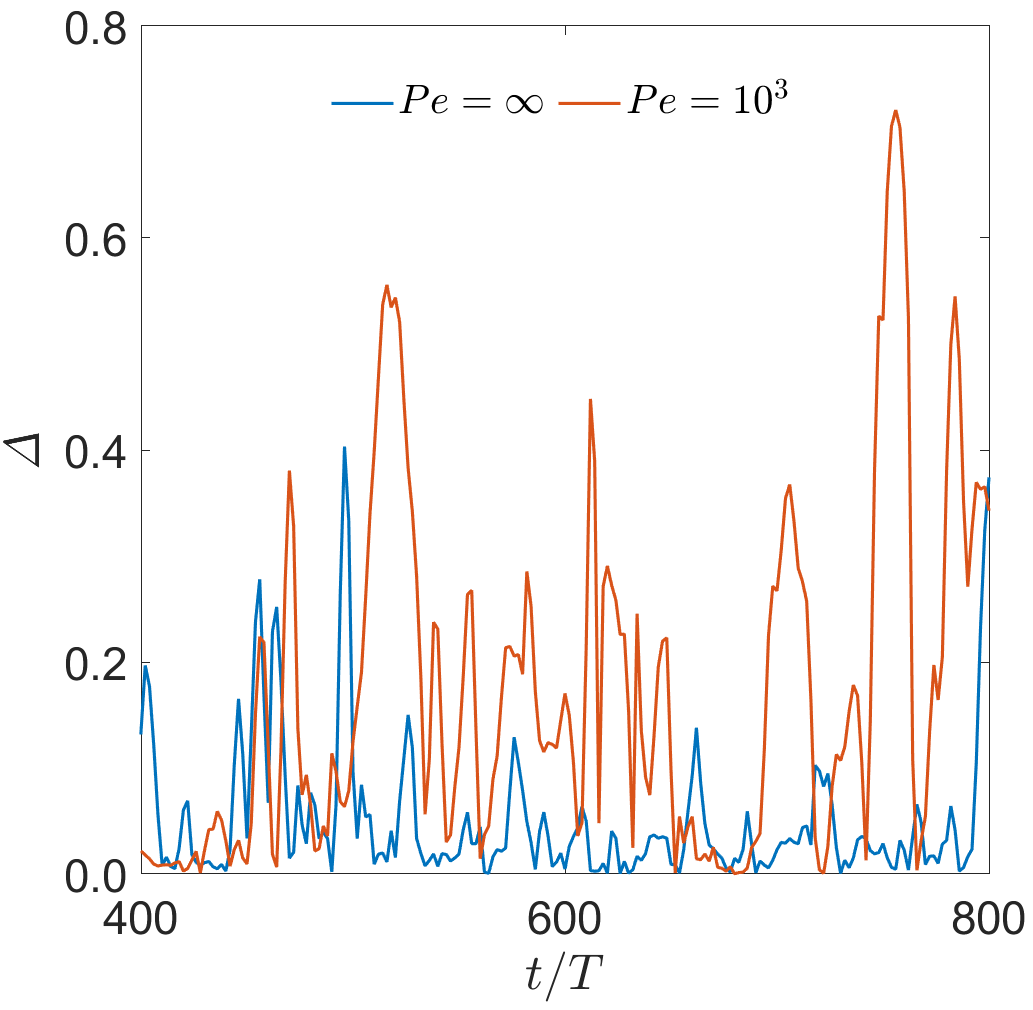}%
\caption{}
\end{subfigure}%
\hfill%
\begin{subfigure}[b]{\textwidth/3}
\includegraphics[width=\textwidth]{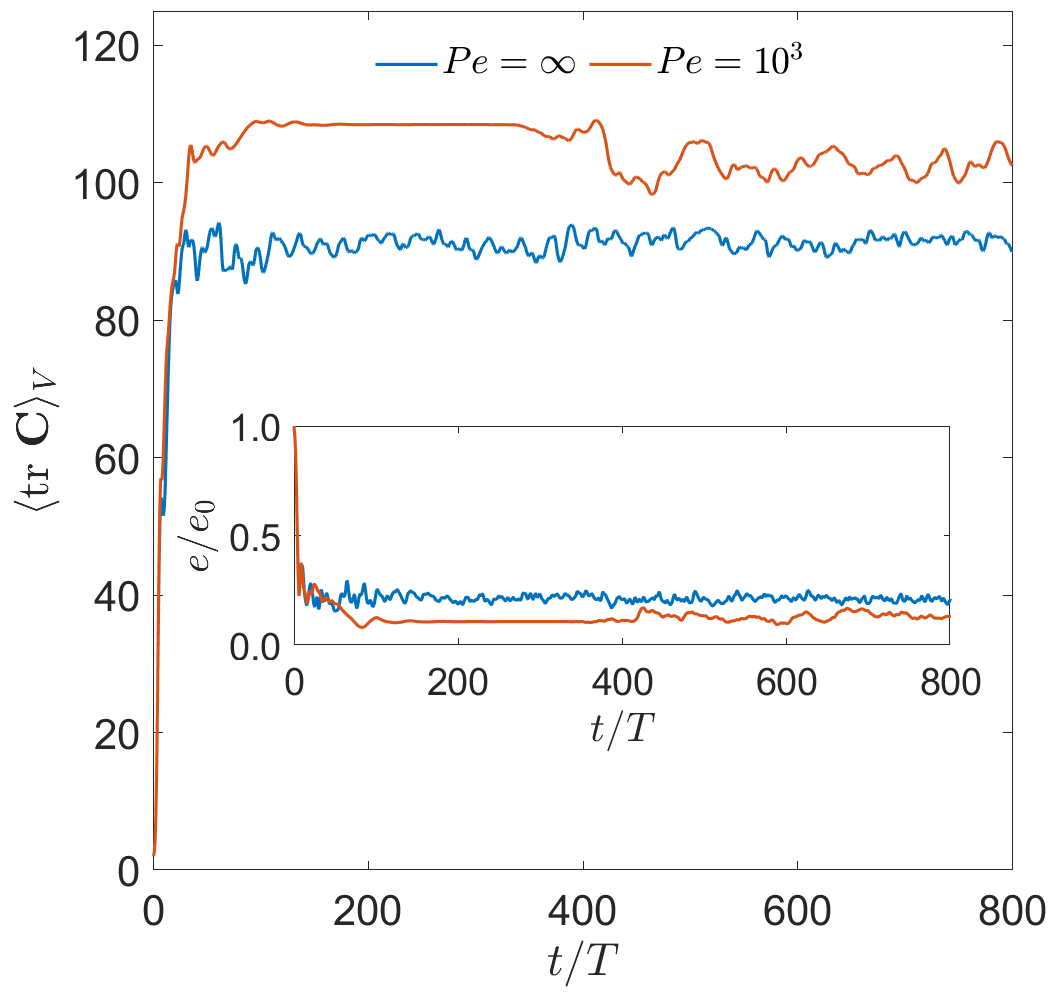}%
\caption{}
\end{subfigure}%
\hfill%
\begin{subfigure}[b]{\textwidth/3}
\includegraphics[width=\textwidth]{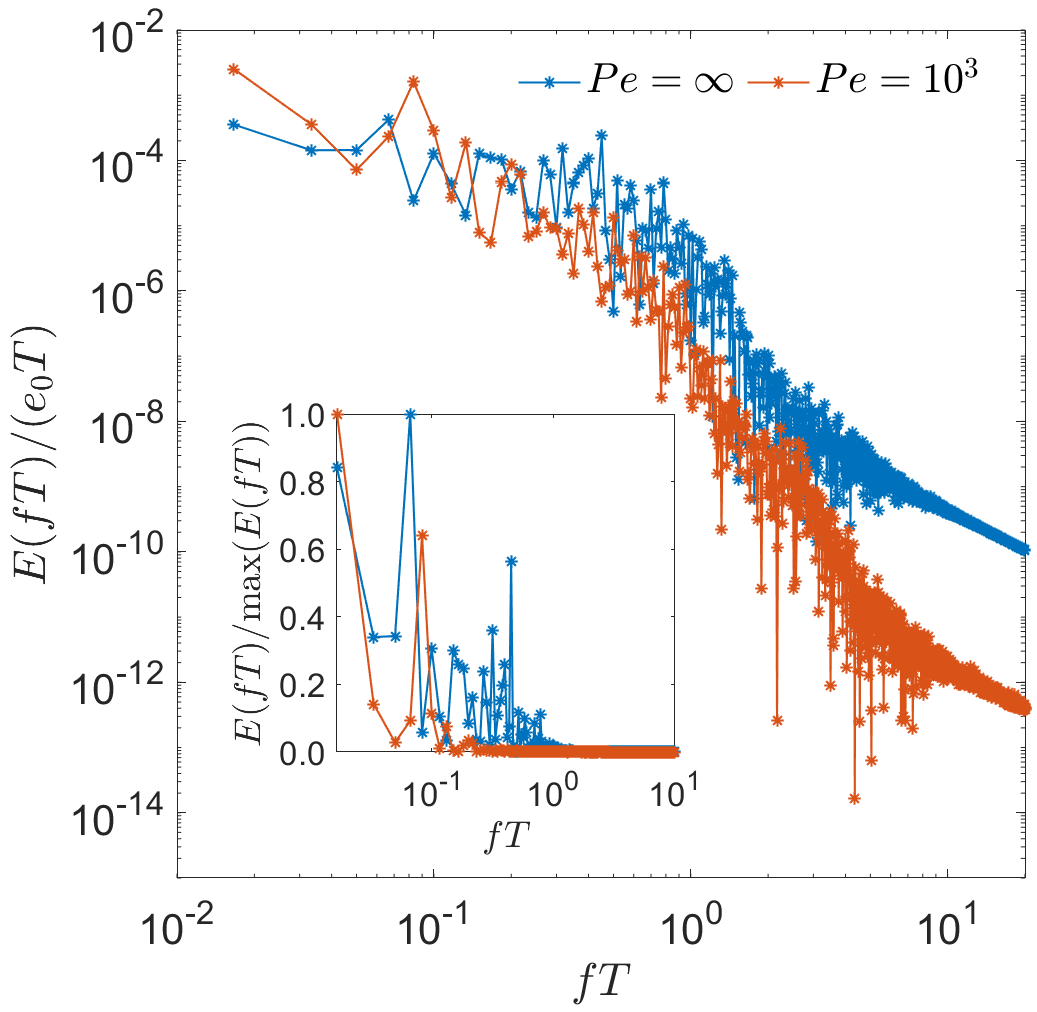}%
\caption{}
\end{subfigure}%
\caption{ (\textit{a}) Time series of  $\varDelta(t)$; (\textit{b}) time series of the space averages of $\trC$ (main panel) and kinetic energy scaled by its Newtonian value $e(t)/e_0$ (inset); (\textit{c}) power spectra of the kinetic-energy fluctuations (the inset shows a semi-log plot of the normalized spectra). Each panel compares the results for the Cholesky-log reformulation of the FENE-P model with $\Pen=\infty$ and $\Pen=10^3$.}
\label{fig:diffusion}
\end{figure}

Artificially enhancing polymer diffusivity, as a means of numerical stabilization, is now well known to produce excessive smearing of the polymeric stress \citep{std18}
and to qualitatively modify the large-scale dynamics in elastic turbulence \citep{gv19}. We now examine whether these artifacts are ameliorated by localizing diffusion to regions where the gradients of $\mathsfbi{C}$ are large. The motivation for trying to retain some form of diffusion is that one can then potentially avoid complex advection schemes, and instead use pseudo-spectral methods that are the workhorse of Newtonian turbulence simulations. Moreover, diffusion also aids in keeping $\mathsfbi{C}$ positive definite \citep{vrbc06}, possibly allowing higher values of $\Wi$ to be attained with moderate spatial resolution. However, these potential benefits are immaterial if localized diffusion ends up modifying the dynamics of elastic turbulence.

We test the effects of local polymer-stress diffusion using the variable diffusivity proposed by \cite{dfs22} (see \S~\ref{sec:nummethod}), which smoothens the $\mathsfbi{C}$ field only at locations where its derivatives are large. In particular, we examine whether local diffusion modifies the large-scale structure of the polymer conformation field, by comparing simulations of the Cholesky-log decomposition with and without local polymer-stress diffusion. We have ensured that the only difference between the two simulations is the addition of a local diffusion term to \eqref{eq:C}, so that any discrepancy in the dynamics must be due to local diffusion. To facilitate a comparison with the results of \cite{dfs22}, we use the same viscoelastic model (FENE-P) and the same peak diffusivity, which corresponds to $\Pen=10^3$. 

Figure~\ref{fig:snapshots-diffusion} compares snapshots of $\trC$ which, along with the associated animations in supplementary \href{https://math.unice.fr/~vincenzi/Movie3.mp4}{movie 3}, show that local diffusion produces misshapen vortical cells and a strong deviation from the forcing pattern. This observation is corroborated by the temporal evolution of $\varDelta(t)$ (see \eqref{eq:Delta}) in the stationary state, depicted by figure~\ref{fig:diffusion}\textit{a}; we see significantly larger fluctuations for $\Pen=10^3$ than for $\Pen=\infty$. This spurious behaviour of the vortical cells in the presence of local stress diffusion is similar to that found by \cite{dfs22}. 

Noticeable differences are also found in the space-averaged dynamics of polymer stretching, presented in figure~\ref{fig:diffusion}\textit{b}. For $\Pen=\infty$, the chaotic regime begins soon after the polymers are stretched out. In contrast, for $\Pen=10^3$, the average extension of polymers is higher and chaotic fluctuations develop only after a long interval of time during which the solution is in a quiescent state. Moreover, once the stationary state is eventually attained, the fluctuations are larger and slower for $\Pen=10^3$.
These differences are naturally reflected in the dynamics of the flow (inset of figure~\ref{fig:diffusion}\textit{b}), which for $\Pen=10^3$ has a lower average kinetic energy (due to stronger polymer stretching and feedback) that fluctuates with larger and slower oscillations. This contrast in temporal fluctuations is evidenced by the power spectra of the space-averaged kinetic-energy, presented in figure~\ref{fig:diffusion}\textit{c}. Local diffusion causes the temporal dynamics to be dominated by a few low-frequency modes, while a much wider set of modes is active in the absence of diffusion (compare the normalized spectra in the inset of figure~\ref{fig:diffusion}\textit{c}).

We note that the spurious initial transient seen in figure~\ref{fig:diffusion}(\textit{b}) for $\Pen = 10^3$ is not the same as the transient observed in the simulations of \cite{dfs22} (also with  $\Pen = 10^3$). Rather, their supplemental movie, for the case of cellular forcing, shows that the stationary chaotic regime is preceded by an initial period, of about $250T$, during which the cellular structure `shakes' without significant distortion. These differences in initial transients are not unexpected given that some parameter values in our simulations differ from those in \cite{dfs22}. The relevant observation is that local diffusion introduces a prolonged spurious transient in addition to producing significant differences in the stationary dynamics. 

In comparison with global diffusion, the spurious effects of local diffusion were found to be relatively mild \citep{dfs22}. However, the above comparison with a non-diffusive simulation shows that local diffusion still distorts the dynamics significantly. Indeed, the differences between $\Pen=\infty$ and $\Pen=10^3$, described above, parallel those that have been observed for global diffusion \citep{gv19}. This qualitative similarity may be understood by recognizing that the two forms of diffusion---local and global---act similarly in regions of the flow where the polymer stress is large and localized, regions which in fact are subject to strong polymer feedback and therefore drive elastic instabilities and dominate the chaotic dynamics.

In conclusion, artificially enhanced polymer-stress diffusion introduces spurious effects in the large-scale dynamics of elastic turbulence, even when the diffusion is local. It is therefore preferable to avoid using artificial diffusion of any form and instead achieve numerical stability by a specialized and accurate treatment of the advection term in the hyperbolic constitutive equation.

\section{Effect of spatial resolution}
\label{sect:resolution}

We have seen, in \S\ref{sect:decomposition}, that the $\mathsfbi{C}$ field develops extremely large gradients and contains thin filamentary zones of large polymer stretching (see figure~\ref{fig:snapshots}). Under these high-$\Wi$ conditions, shock-capturing advection schemes, like the Kurganov-Tadmor method used here, are essential for maintaining numerical stability because they prevent the gradients from steepening indefinitely. The width of the shock-like large stretching zones is cutoff at the grid scale and not allowed to decrease further. Importantly, the Kurganov-Tadmor method advects the stretching zones accurately and, unlike artificial diffusion, limits the large gradients without spreading the polymeric stress. 

\begin{figure}
\centering
\begin{subfigure}[b]{\textwidth/2}
\centering
\includegraphics[width=\textwidth]{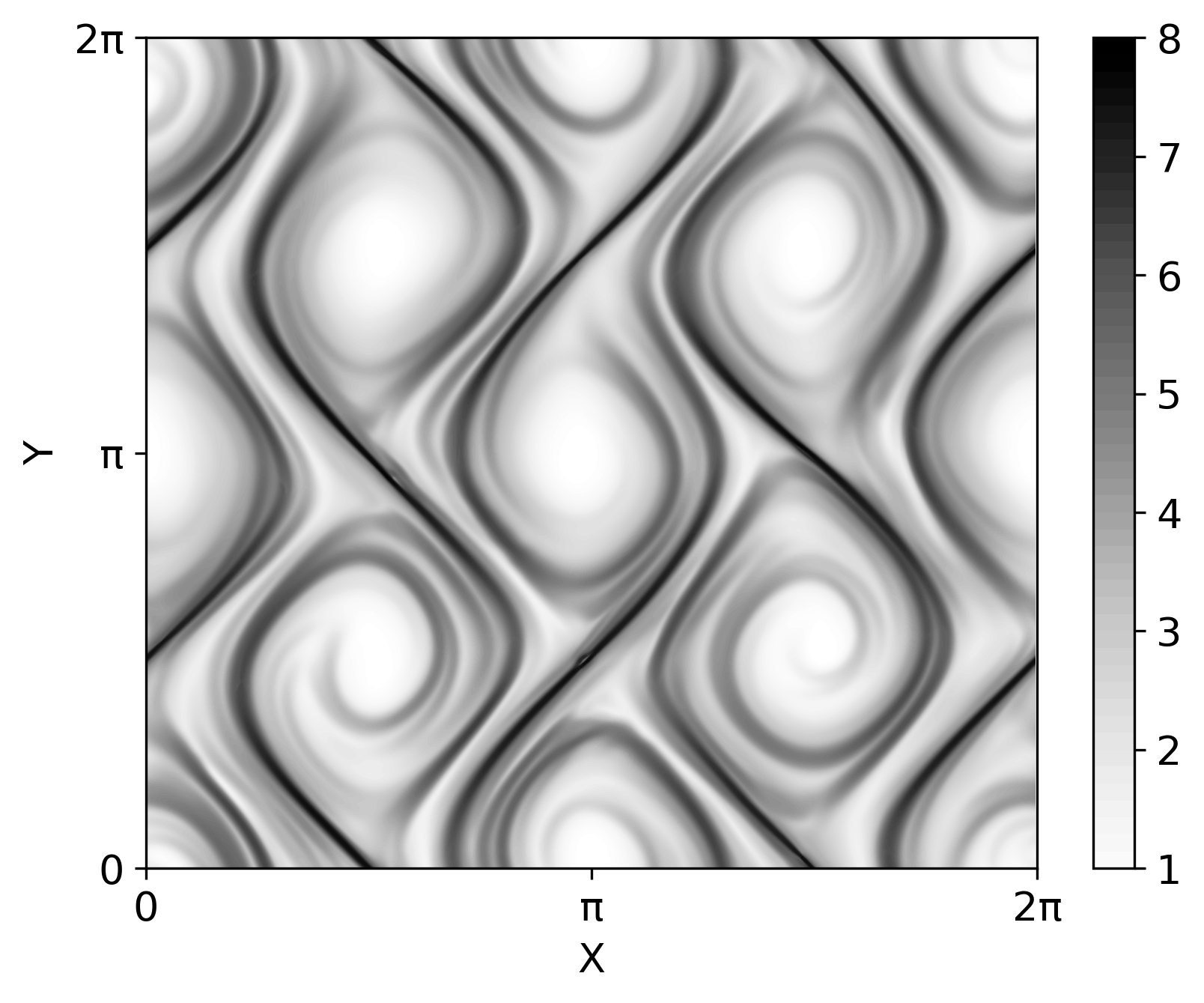}%
\label{OK}
\caption{}
\end{subfigure}%
\hfill%
\begin{subfigure}[b]{\textwidth/2}
\centering
\includegraphics[width=\textwidth]{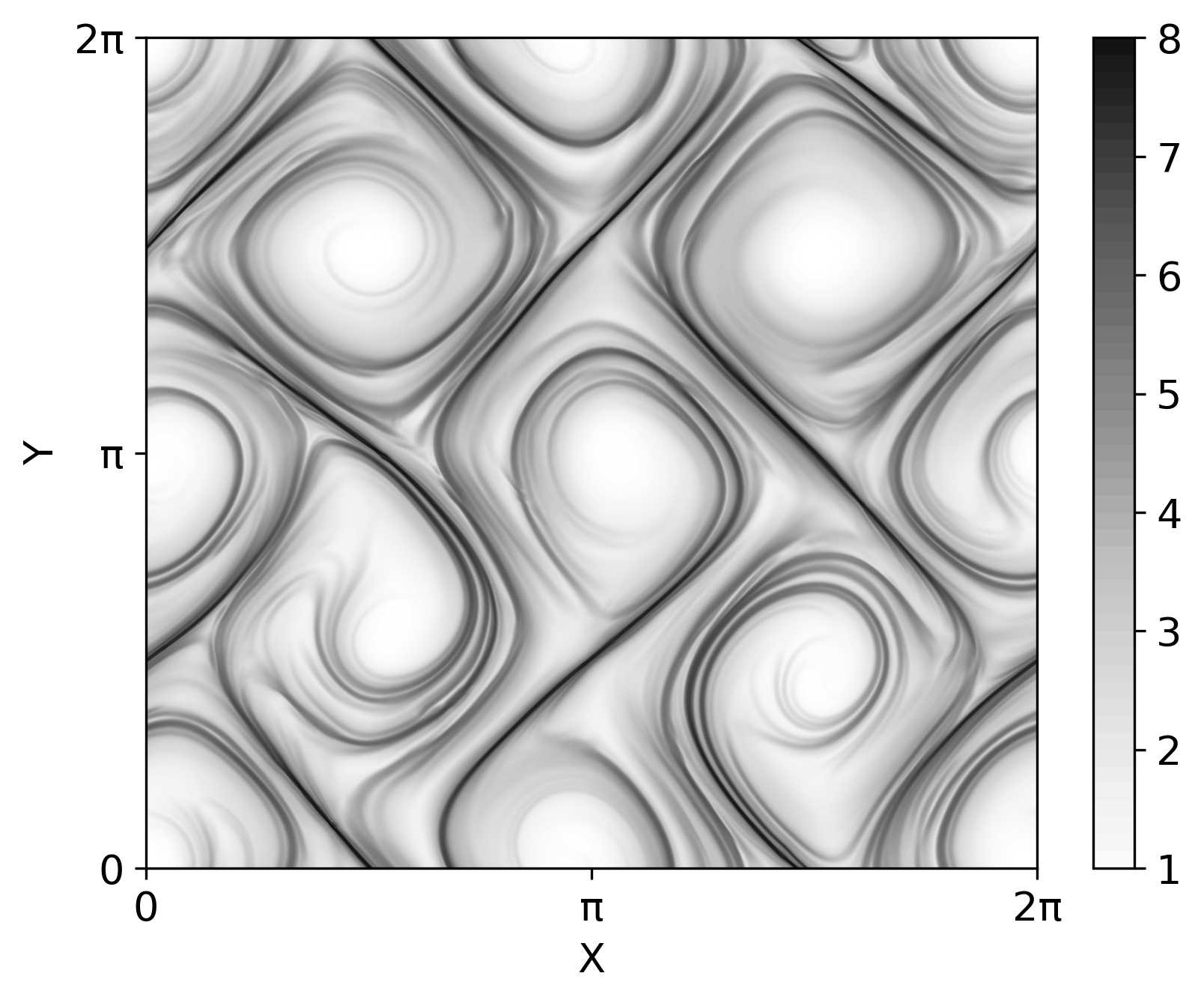}%
\caption{}
\end{subfigure}%

\bigskip
\begin{subfigure}[b]{\textwidth/2}
\centering
\includegraphics[width=.93\textwidth]{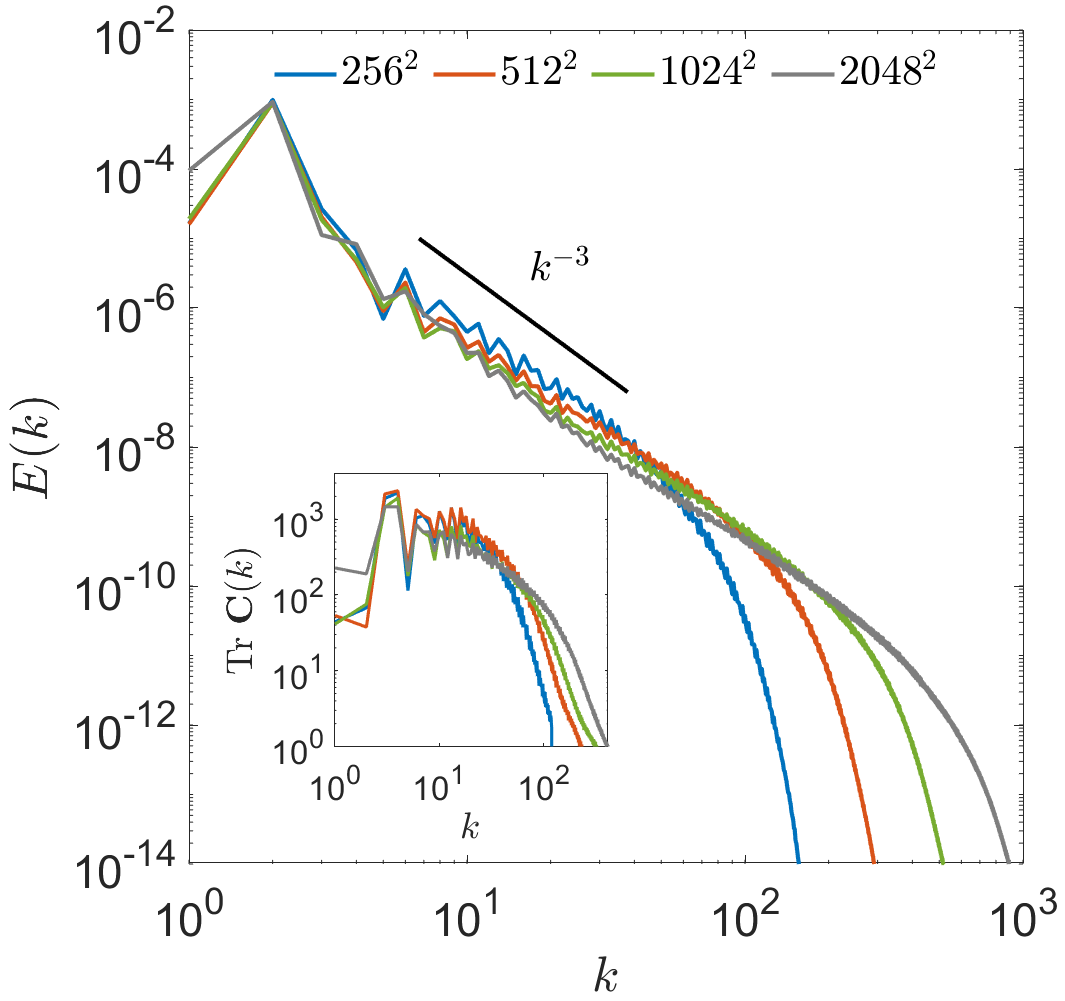}%
\label{OK}
\caption{}
\end{subfigure}%
\hfill%
\begin{subfigure}[b]{\textwidth/2}
\centering
\includegraphics[width=0.89\textwidth]{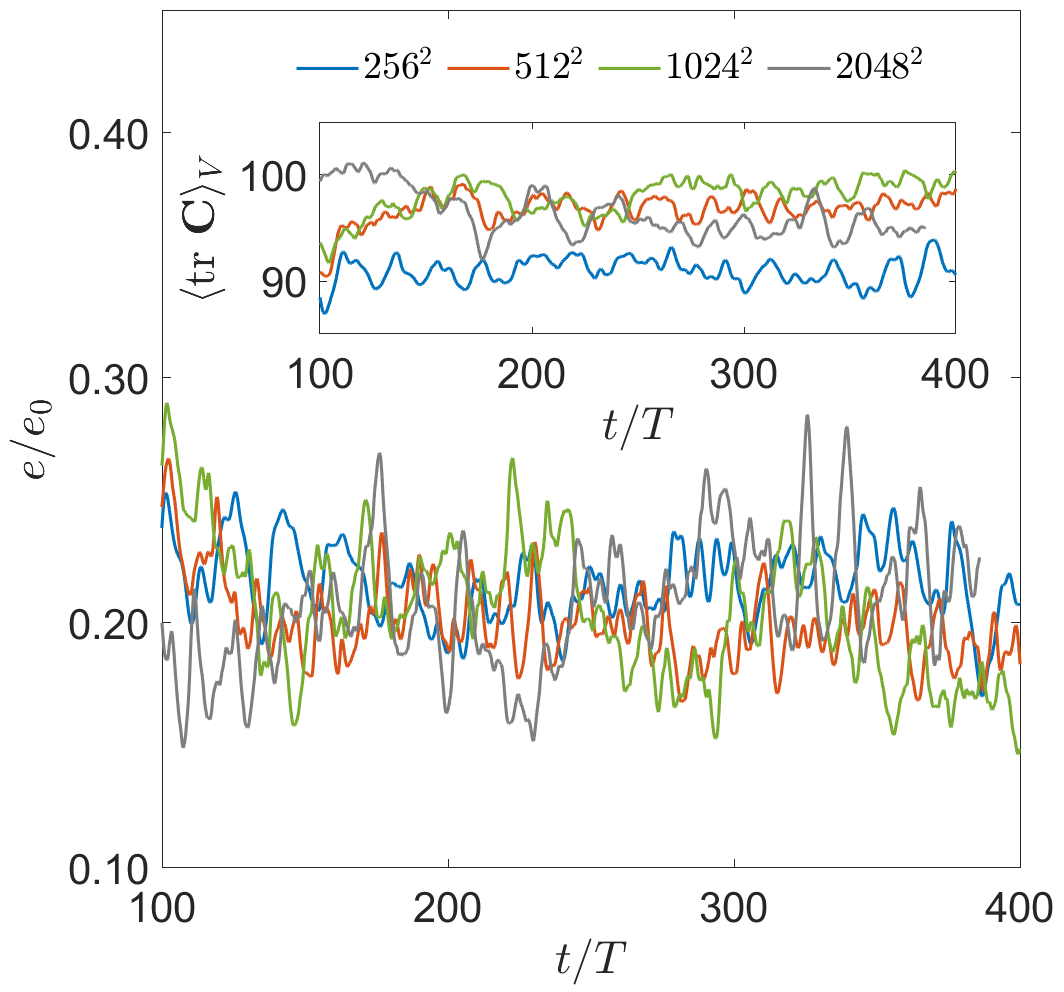}%
\caption{}
\end{subfigure}%
\caption{Representative snapshots of the logarithm of $\trC$ for resolutions (\textit{a})  $512^2$ and (\textit{b})~$2048^2$ (the $256^2$ version is shown in figure~\ref{fig:snapshots-diffusion}\textit{a}); (\textit{c})~energy spectra $E(K)$ and the spectra of $\trC$ (inset) for different resolutions; 
(\textit{d})~time series of the space averages of kinetic energy scaled by its Newtonian value $e/e_0$ (main panel) and $\trC$ (inset) for different resolutions. All simulations have been performed using the Cholesky-log reformulation of the FENE-P model with $\Pen=\infty$.}
\label{fig:resolution}
\end{figure}

While the width of the stretching zones will keep decreasing with increasing spatial resolution, the key dynamical features of the flow should converge. We check if this is true by comparing Cholesky-log simulations of the FENE-P model at spatial resolutions of $256^2$, $512^2$, $1024^2$, and $2048^2$. Snapshots of $\trC$ from $512^2$ and $2048^2$ simulations are presented in figures~\ref{fig:resolution}\textit{(a)} and \ref{fig:resolution}\textit{(b)} respectively (the $256^2$ version is shown in figure~\ref{fig:snapshots-diffusion}\textit{a}). We see that the large-scale cellular structures are very similar at all resolutions, though the stretching zones in the $2048^2$ simulation are much thinner and more refined.

The sharpening of stretching zones with increasing resolution is reflected in the spatial spectrum of $\trC$ (inset of figure~\ref{fig:resolution}\textit{c}), which preserves its form but extends to larger wavenumbers. The same is true of the kinetic energy spectrum $E(k)$ (main panel of figure~\ref{fig:resolution}\textit{c}), which exhibits a power-law over a limited range followed by a tail that extends to higher wavenumbers as the resolution is increased. The mean polymer stretching (inset of figure~\ref{fig:resolution}\textit{d}) increases initially when the resolution is doubled from $256^2$ to $512^2$, likely because larger gradients of $\mathsfbi{C}$ are permitted. However, the mean stretching does not increase further despite a subsequent four-fold increase in the resolution, suggesting that the extent of polymer stretching is ultimately limited by the chaotic dynamics of the flow. The mean kinetic energy (main panel of figure~\ref{fig:resolution}\textit{d}) does not change appreciably with resolution.

On the whole, the considerable thinning of the stretching zones, accompanying the increase of the resolution from $256^2$ to $2048^2$, does not alter the qualitative dynamics of the flow. There are, however, some subtle differences in the $2048^2$ simulation: the spectral energy content in the $k=1$ mode is higher in both spectra (figure~\ref{fig:resolution}\textit{c}), and the kinetic energy exhibits slightly larger fluctuations (figure~\ref{fig:resolution}\textit{d}). A systematic investigation of these variations would require simulations at even higher resolutions which are beyond the scope of this work. Nonetheless, it is clear that, apart from minor quantitative changes, the use of the Kurganov-Tadmor scheme along with the log-Cholesky decomposition enable the key features of the flow to be captured even at a relatively low resolution of $256^2$.

\section{Implications for predicting mixing}
\label{sect:mixing}

Most applications of elastic turbulence are based on its enhanced mixing. We now show that this important property is substantially modified by the erroneous large-scale dynamics that result from using matrix decompositions without a logarithmic transformation (\S~\ref{sect:decomposition}) or as a consequence of local polymer-stress diffusion (\S~\ref{sect:diffusion}).

Consider the transport of a scalar blob that is passively dispersed by the flow. The scalar concentration $\theta(\boldsymbol{x},t)$ satisfies the advection-diffusion equation
\begin{equation}
\partial_t \theta +\boldsymbol{u}\bcdot\bnabla\theta = \kappa_\theta \Delta\theta,
\label{eq:scalar}
\end{equation}
where the scalar diffusivity $\kappa_\theta=10^{-5}$ is associated with a scalar P\'eclet number $\Pen_\theta = 5\times10^3$. Equation \eqref{eq:scalar} is solved using the same finite-difference scheme as that used for the constitutive equation; thanks to its diffusion, though, no special treatment is needed to keep $\theta$ positive. In what follows, we begin the evolution of the scalar field only after the flow has attained stationarity.

A distinguishing feature of the accurate large-scale dynamics is that the cellular structure of the forcing is well-preserved, unlike the erroneous simulations which have distorted vortical cells that constantly change their shape and size (see figure~\ref{fig:snapshots}). Now, if a scalar blob is placed within one of the vortical cells, then its dispersion is bound to be impacted by the robustness of the cellular structure or the lack thereof. We therefore choose an initial configuration in which the scalar is concentrated over a disc of centre $(\upi,\upi)$ and radius $r=0.4$, \textit{i.e.}, a blob is initially placed inside the central vortical cell.

\begin{figure}
\centering
\begin{subfigure}[b]{\textwidth/2}
\centering
\includegraphics[width=.95\textwidth]{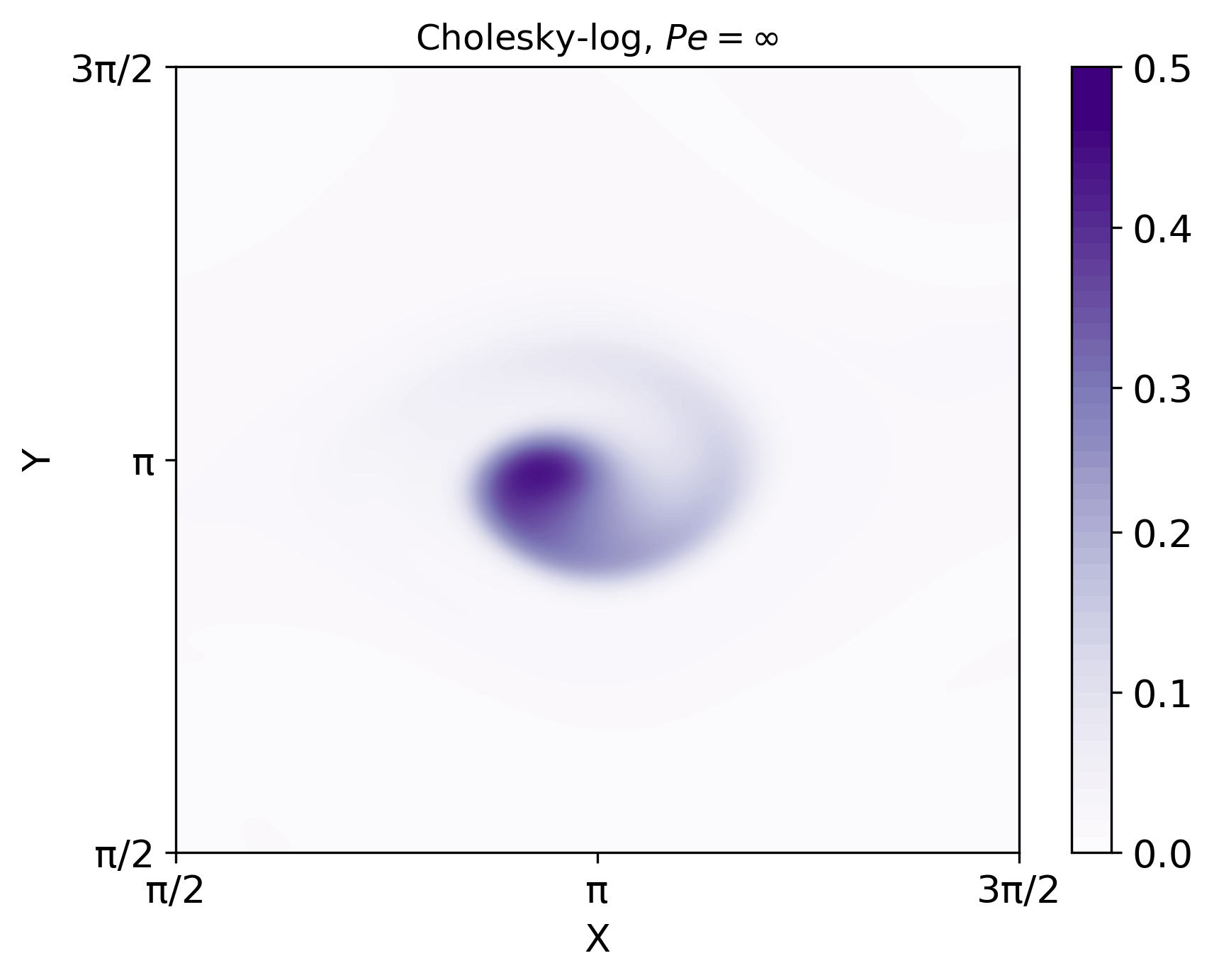}%
\label{OK}
\caption{}
\end{subfigure}%
\hfill%
\begin{subfigure}[b]{\textwidth/2}
\centering
\includegraphics[width=.95\textwidth]{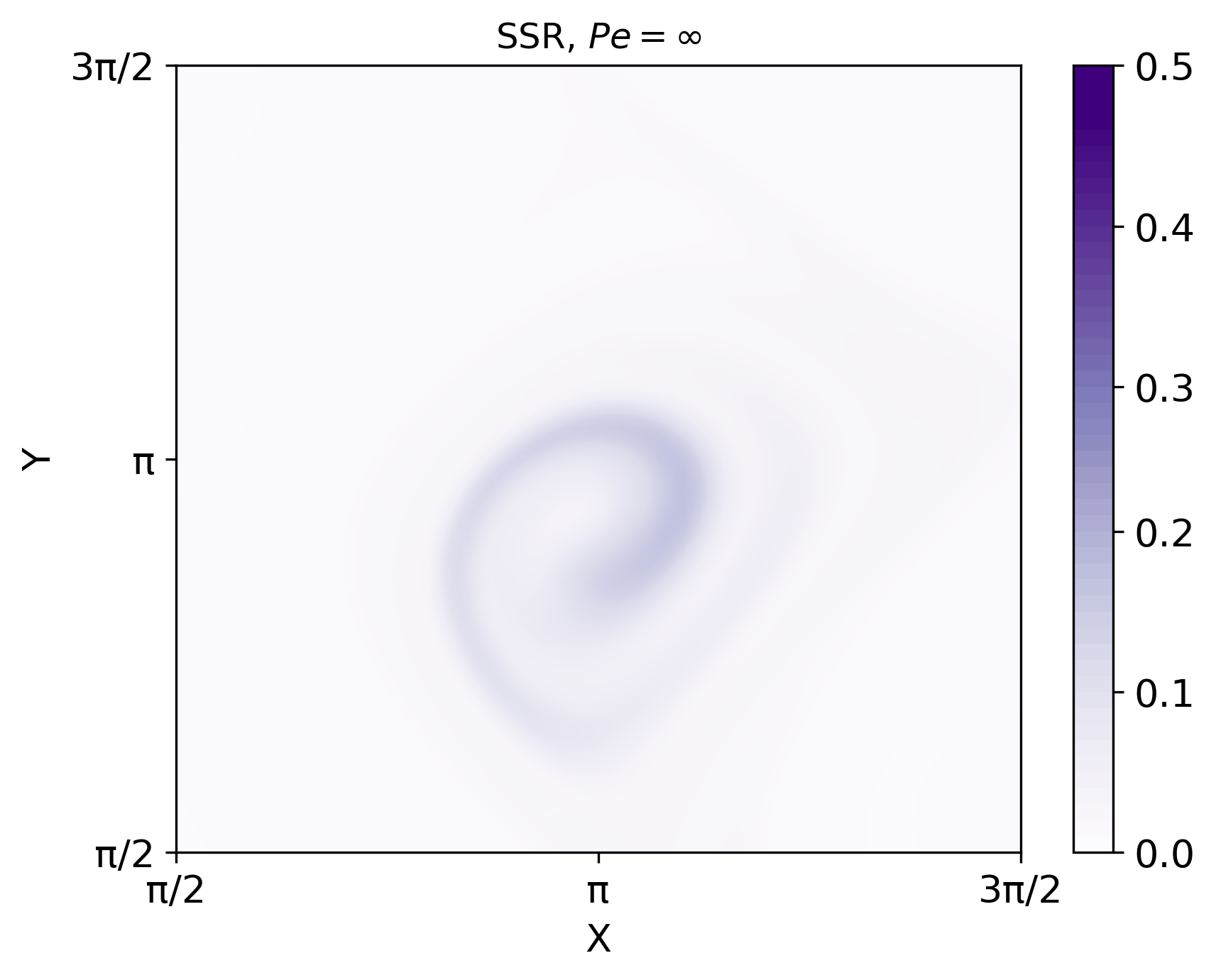}%
\caption{}
\end{subfigure}%
\\
\begin{subfigure}[b]{\textwidth/2}
\centering
\includegraphics[width=.95\textwidth]{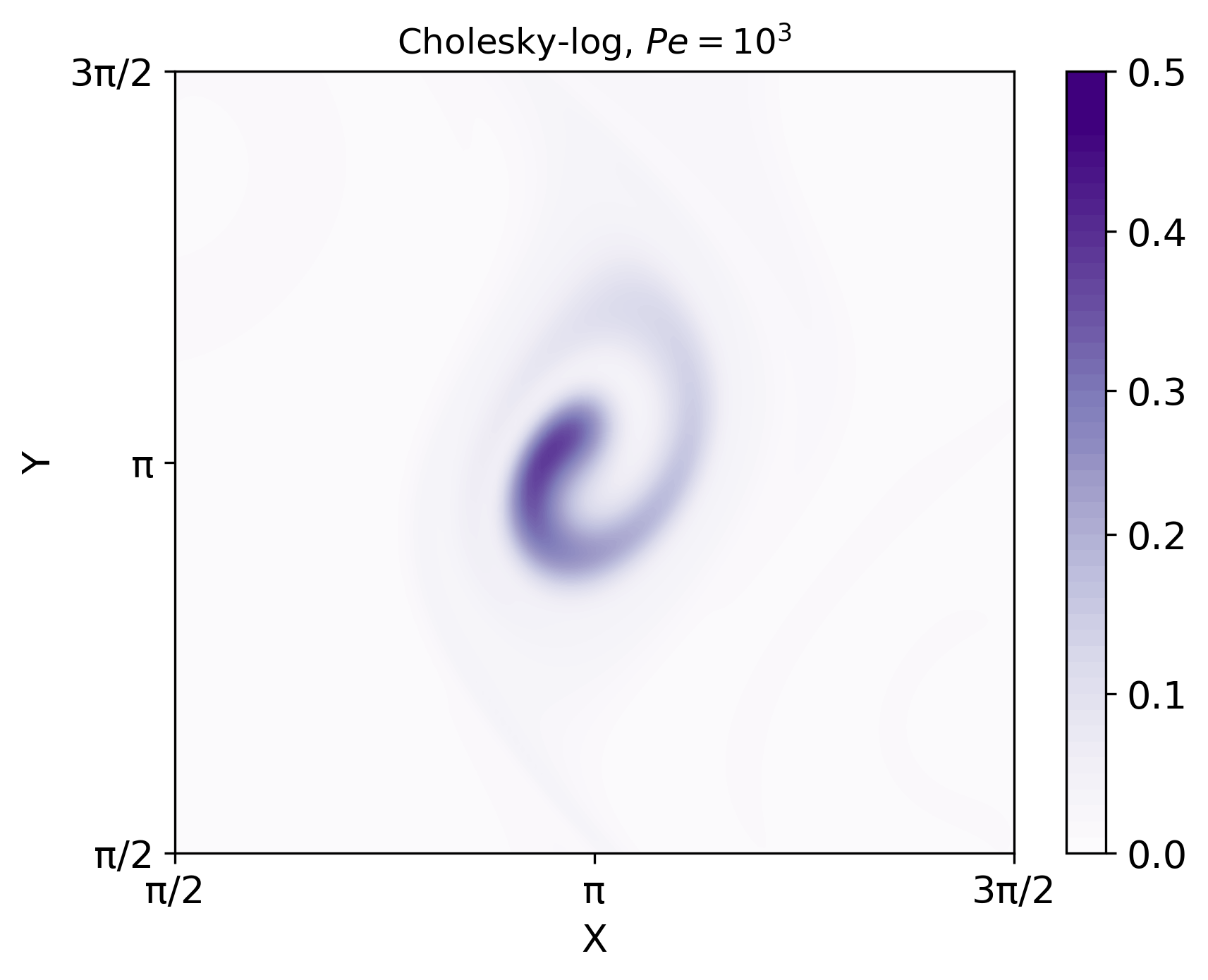}%
\label{OK}
\caption{}
\end{subfigure}%
\hfill%
\begin{subfigure}[b]{\textwidth/2}
\centering
\includegraphics[width=0.75\textwidth]{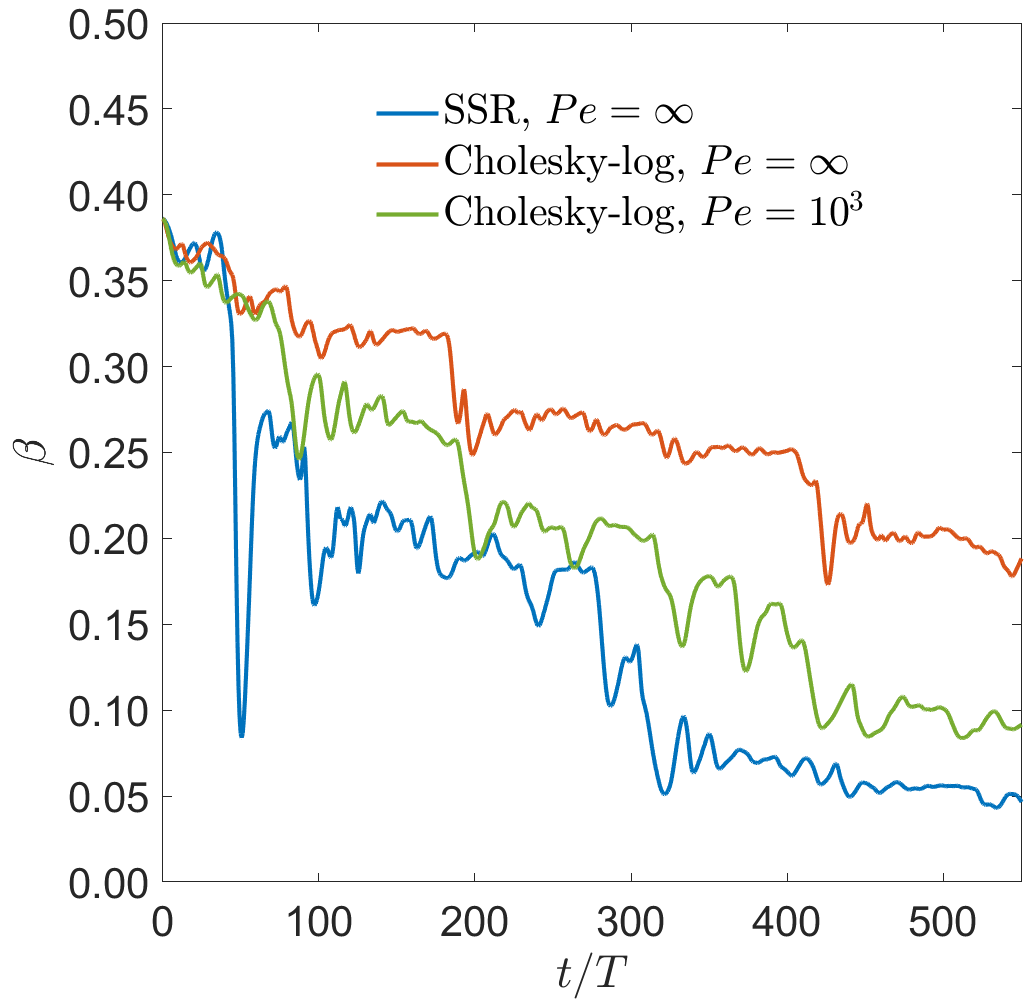}%
\caption{}
\end{subfigure}%
\caption{Snapshots of the scalar concentration $\theta(\boldsymbol{x},t)$ at time $t=180T$ for (\textit{a}) the Cholesky-log decomposition with $\Pen=\infty$, (\textit{b})~the SSR decomposition with $\Pen=\infty$, and (\textit{c}) the Cholesky-log decomposition with $\Pen=10^3$. 
For the corresponding animations, see supplementary \href{https://math.unice.fr/~vincenzi/Movie4.mp4}{movie 4}. 
The snapshots have been zoomed over the region $[\upi/2,3\upi/2]^2$ to emphasize the central cell; the movie shows the entire domain $V$. 
Panel~(\textit{d}) shows the mixing diagnostic $\beta(t)$ for the three simulations. All simulations are based on the FENE-P model.}
\label{fig:mixing}
\end{figure}

Snapshots of the scalar field, after it has evolved for $180T$, are presented in figure~\ref{fig:mixing}, for simulations of the FENE-P model using 
(a) the Cholesky-log decomposition with $\Pen=\infty$, (b) the SSR decomposition with $\Pen=\infty$, and (c) the Choleksy-log decomposition with $\Pen=10^3$.
The associated animations are available in supplementary \href{https://math.unice.fr/~vincenzi/Movie4.mp4}{movie 4}. Clearly, the extent to which the scalar has migrated out of the initial cell is much greater in case of the SSR decomposition than for the Cholesky-log decomposition (compare figures~\ref{fig:mixing}\textit{b} and~\ref{fig:mixing}\textit{a}). The case with local polymer-stress diffusion also shows more dispersal than that without (compare figures~\ref{fig:mixing}\textit{c} and~\ref{fig:mixing}\textit{a}). These differences are obviously due to the spurious distortion of the vortical cells which arises due to either numerical inaccuracies in the SSR simulation or artificial spreading of polymeric stress in the presence of local stress-diffusion.

To quantify the dispersion of the scalar blob, we consider the quantity $\beta(t)=\langle\theta(t)\rangle_D - \langle\theta(0)\rangle_V$, where $\langle\cdot\rangle_V=(2\upi)^{-2}\int_V \cdot \,\mathrm{d}\boldsymbol{x}$ is the average over the entire domain and $\langle\cdot\rangle_D=(4\upi r^2)^{-1}\int_D \cdot \,\mathrm{d}\boldsymbol{x}$ is the average over a disc $D=\{(x,y) \in V \ : \ (x-\upi)^2+(y-\upi)^2 \leq 2r\}$, which has twice the radius of the initial blob. When the scalar is completely mixed, then the average over the disc $D$ will equal the average over the entire domain, which being preserved by \eqref{eq:scalar} will remain at its initial value of $\langle\theta(0)\rangle_V$. Thus, $\beta(t)$ will ultimately tend to zero as the scalar field becomes uniform at long times. Of course, $\beta(t)$ need not decrease monotonically because the scalar can be transported in and out of the disc $D$. In fact, we have chosen such a diagnostic because it may reveal differences in the early dispersion process that would be averaged out by quantities like the variance which average over the entire domain. 

Figure~\ref{fig:mixing}(\textit{d}) compares the time trace of $\beta(t)$ for the three simulations and confirms that the dispersion is slowest for the Cholesky-log decomposition. These time traces are characterized by sudden drops, which correspond to instances when a significant fraction of the scalar leaks out of the central cell. Such instances are few and far between in the Cholesky-log simulation ($\Pen = \infty$), wherein the cells are just mildly perturbed, but quite frequent in the other two simulations, wherein the cells are strongly distorted. 

These results demonstrate that even though the different solution methods all give rise to a chaotic flow, the discrepancies in their detailed dynamics, uncovered in this study, have a significant impact on the mixing properties of the flow.

\section{Summary and conclusions}
\label{sect:conclusions}

The simulation of elastic turbulence presents difficulties associated with the advection of the polymer stress.
The absence of a diffusive term (or one that can be resolved in practice) in the constitutive equation leads to the formation of strong gradients in the polymer conformation tensor. If these are not resolved and advected accurately, the mathematical properties of the conformation tensor may be violated, and ultimately the entire dynamics of the polymer solution may be misrepresented.

In the literature, a variety of strategies have been proposed to overcome the aforementioned numerical difficulties. The aim of this study was to identify the combination of numerical methods that are most suitable for the simulation of elastic turbulence. Our conclusion is that preference should be given to those methods that involve a logarithmic reformulation of the constitutive equation, incorporate a shock-capturing scheme for the advection of the conformation tensor, and do not make use of artificial polymer-stress diffusion, even if localized.

In high-strain regions, the use of a logarithmic reformulation is known to reduce the numerical errors associated with exponential stretching.
We have shown that, in elastic turbulence simulations, this is important to preserve the properties of the polymer conformation tensor and reproduce the large-scale dynamics faithfully. 
Although we have focused on the Cholesky-log decomposition, we expect similar conclusions to apply to the log-conformation representation, since the basic principles behind the two approaches are analogous. A point in favour of using the log-conformation method is that it would facilitate the calculation of a recently-introduced diagnostic, the log-Euclidean mean conformation tensor \citep{hameduddin2019}, that can aid in understanding and modelling viscoelastic turbulence.

Prior to this work, it was known that applying artificial polymer-stress diffusion locally, rather than uniformly in space and time, attenuates the artifacts that arise due to excessive smoothing of the polymer-stress field. It was unclear, however, whether the effects of localized stress diffusion are truly negligible or can still modify the large-scale dynamics and, if so, to what extent. We have shown here that since local diffusion intervenes at locations where the polymer stress is large and localized, that is, precisely where the polymer feedback is strong, it ends up having a significant effect on the dynamics of the flow. Indeed, the redistribution of the largest stresses destabilizes the large flow structures in a similar way, if not to the same extent, as global artificial diffusion.
The use of artificial polymer-stress diffusion, of any form, in elastic turbulence simulations should therefore be discouraged. (We have also tried using global and local hyperdiffusion with a fourth order Laplacian, only to find once again that the large-scale dynamics change substantially). This result leads to a preference for finite difference methods over pseudo-spectral methods. Indeed, while the latter generally require some form of small-scale dissipation for stability, the former can more easily incorporate shock-capturing schemes to resolve the discontinuities that form in the polymer conformation field.

Our study also explains some observations of \cite{btrd11},  who used the SSR decomposition to simulate two-dimensional elastic turbulence, driven by a four-roll mill forcing. 
Their simulations displayed a strong destabilization of the cellular symmetry of the forcing---a characteristic artifact of global diffusion \citep{gv19,dfs22}---even though they had abstained from adding stress diffusion to the constitutive equation. 
This loss of symmetry can now be explained as a consequence of using the SSR decomposition which, as we have shown, gives rise to numerical inaccuracies that modify the large-scale dynamics in a manner similar to global or local artificial diffusion. Furthermore, \cite{btrd11} do not use a shock-capturing advection scheme, but instead use a pseudo-spectral method. And though they implement a smooth filter to damp high-wavenumber modes (an operation that is similar in spirit to using local hyperdiffusivity) their simulations suffer from spurious spatial oscillations in the vicinity of sharp gradients in the conformation tensor. 


The recommended combination of a logarithm-based reformation of the conformation tensor along with a non-diffusive shock-capturing advection scheme allows us to predict the large-scale chaotic dynamics of the flow rather inexpensively. Indeed, the results of our simulations, using Cholesky-log and Kurganov-Tadmor, remain qualitatively unchanged with regard to the large scales and their spatial structure from a resolution of $256^2$ to $2048^2$, though the quasi-discontinuous stretching zones become ever sharper and more refined with increasing resolution. At the same time, the erroneous large-scale dynamics, obtained when using the decomposition without a logarithmic transformation or when using local artificial diffusion, were not cured by increasing the resolution. Thus, a high resolution while preferable is not as important a factor for elastic turbulence simulations as a log-based reformulation of the conformation tensor, the absence of artificial diffusion, and an accurate advection scheme.


Our analysis has also emphasized a mathematical lower bound on the determinant of the conformation tensor that has apparently been disregarded in numerical simulations of viscoelastic flows. In the Oldroyd-B model, the determinant of the conformation tensor must not only be positive, as required by the positive-definite nature of the tensor, but also remain greater than unity at all times (after an initial transient, in case it starts out below unity). Thus, ensuring positive definiteness is not sufficient for numerical accuracy. As we have seen, even when the determinant of the conformation tensor stays positive, thanks to a suitable matrix decomposition, its value can fall below unity, which in case of the Oldroyd-B model signifies numerical errors that are significant enough to modify the large-scale dynamics of the polymer solution. Since the bound on the determinant of the conformation tensor applies only to the Oldroyd-B model, we recommend testing new numerical schemes or simulation codes first with the Oldroyd-B model, to ensure that the bound is satisfied, before applying them to other viscoelastic models.

While we have focused on elastic turbulence and the corresponding low-$\Rey$ limit, the bound on the determinant of the conformation tensor is valid for high-$\Rey$ flows as well. Preliminary simulations of high-$\Rey$ viscoelastic turbulent flows have shown that, just as for the low-$\Rey$ limit, simulations using the Cholesky-log decomposition preserve the bound whereas those using the SSR decomposition violate it \citep{sumithra}.  However, the impact of the numerical errors on the dynamics of elasto-inertial turbulence at moderate-$\Rey$ and drag-reduction at high-$\Rey$ remain to be investigated. Nonetheless, it is clear that a logarithmic reformulation of the conformation tensor remains preferable even when $\Rey$ is not small.

Finally, through the study of the dispersion of a scalar blob, we have shown that errors in computing the large-scale structures of the flow translate into erroneous predictions of mixing. Therefore, an accurate numerical method that resolves the advection of polymer stresses is important for studying not only details of the chaotic dynamics but also applications of elastic turbulence.\\


\small{\textbf{Acknowledgements.} {J.R.P. and D.V. acknowledge their Associateships with the International Centre for Theoretical Sciences (ICTS), Tata Institute of Fundamental Research, Bangalore,
India. S.R.Y. and D.V. thank the OPAL infrastructure and the Center for High-Performance Computing of Universit\'e C\^ote d’Azur, and the ICTS, Bangalore, for computational resources.  Simulations were also performed on the IIT Bombay workstations \textit{Aragorn} and \textit{Gandalf} (procured through grant SRG/2021/001185). A.G. would like to thank NSM facility \textit{PARAM-SHIVAY} at IIT-BHU and \textit{PARAM-SEVA} at IIT Hyderabad, and mini-HPC \textit{Kanad} at Department of Physics, IIT Hyderabad}}\\

\small{\textbf{Funding.} {This work was supported by CNRS  (S.R.Y., 80 | Prime program);
Agence Nationale de la Recherche (S.R.Y. and D.V., grant no. ANR-15-IDEX-01);
IRCC, IIT Bombay (J.R.P., seed grant); DST-SERB (J.R.P., grant no. SRG/2021/001185; A.G., grant no. MTR/2022/000232);
F\'ed\'eration de Recherche Wolfgang Doeblin (J.R.P.); DST (A.G., grant no. DST/NSM/R\&D\_HPC\_Applications/2021/05 and grant no. SR/FST/PSI-215/2016);
the Indo--French Centre for Applied Mathematics IFCAM (S.R.Y., J.R.P. and D.V.); and
the Indo–French Centre for the Promotion of Advanced Scientific Research (IFCPAR / CEFIPRA) (J.R.P. and D.V., project no. 6704-A).
}\\

\small{\textbf{Author ORCHID.} {S.~R.~Yerasi, https://orcid.org/0000-0003-3095-959X; J.~R.~Picardo, https://orcid.org/0000-0002-9227-5516; A.~Gupta, https://orcid.org/0000-0002-7335-0584; D.~Vincenzi, https://orcid.org/0000-0003-3332-3802}

\appendix

\section{Decompositions of the polymer conformation tensor}
\label{app:matrix}

\subsection{Cholesky-log decomposition}

In three dimensions, the evolutions equations for  the Cholesky factor of $\mathsfbi{J}=f(r)\mathsfbi{C}$ can be found in \citet{vc03} (a misprint in the equation for $\mathsfi{L}_{32}$ is corrected in \citealt{pmp06}).
For $d=2$,  the equations simplify to (see \citealt{gpp15}; note that a sign error in the equation for $\mathsfi{L}_{22}$ is corrected below)
\begin{subequations}
\label{eq:cholesky}
\begin{align}
\label{l11cholesky}
\frac{D \mathsfi{L}_{11}}{D t} &= (\partial_x u_x) \mathsfi{L}_{11}+(\partial_y u_x) \mathsfi{L}_{21} + \frac{1}{2}\left[\frac{p}{\mathsfi{L}_{11}}+(q-p)\mathsfi{L}_{11}\right],
\\ 
\label{l12cholesky}
\frac{D \mathsfi{L}_{21}}{D t} &= (\partial_x u_y) \mathsfi{L}_{11}+(\partial_y u_y) \mathsfi{L}_{21} + (\partial_y u_x)\frac{\mathsfi{L}_{22}^2}{\mathsfi{L}_{11}} +\frac{1}{2}\left[(q-p)\mathsfi{L}_{21}-\frac{p\mathsfi{L}_{21}}{\mathsfi{L}_{11}^2}\right],
\\
\label{l22cholesky}
\frac{D \mathsfi{L}_{22}}{D t} &= (\partial_y u_y) \mathsfi{L}_{22}-(\partial_y u_x) \frac{\mathsfi{L}_{21}\mathsfi{L}_{22}}{\mathsfi{L}_{11}} +  \frac{1}{2}\left[(q-p)\mathsfi{L}_{22}+\frac{p}{\mathsfi{L}_{22}}+\frac{p \mathsfi{L}_{21}^2}{\mathsfi{L}_{11}^2 \mathsfi{L}_{22}}\right],
\end{align}
\end{subequations}
where
\begin{align}
p = \frac{b -2 + j^2}{b \tau_p }, 
&\quad
q = \left[\frac{h}{b -2} - \frac{(b -2 + j^2)(j^2-2)}{b (b-2)\tau_p}\right], 
\\[3mm]
j^2=\operatorname{tr}\mathsfbi{J}, 
& \quad
h = \operatorname{tr}[\mathsfbi{J}\bcdot\bnabla\boldsymbol{u}+(\bnabla \boldsymbol{u})^\top\bcdot \mathsfbi{J}].
\end{align}
To enforce the positivity of the diagonal elements of $\mathsfbi{L}$, \cite{vc03} evolved $\tilde{\mathsfi{L}}_{ii}=\ln\mathsfi{L}_{ii}$ and calculated $\mathsfi{L}_{ii}$ by taking the exponential of  $\tilde{\mathsfi{L}}_{ii}$.
The evolution equations for  $\tilde{\mathsfi{L}}_{ii}$ are:
\begin{subequations}
\begin{align}
\label{l11tilda}
\frac{D \tilde{\mathsfi{L}}_{11}}{D t}  & =  \partial_x u_x +(\partial_y u_x) \mathsfi{L}_{21}\exp\big(-\tilde{\mathsfi{L}}_{11}\big) + \frac{1}{2}\left[p\exp\big(-2\tilde{\mathsfi{L}}_{11}\big)+q-p\right], 
\\
\label{l22tilda}
\frac{D \tilde{\mathsfi{L}}_{22}}{D t}
& =   \partial_y u_y -(\partial_y u_x) \mathsfi{L}_{21}\exp\big(-\tilde{\mathsfi{L}}_{11}\big) 
\\
& \quad +    \frac{1}{2}\left\{p\exp\big(-2\tilde{\mathsfi{L}}_{22}\big)+p\mathsfi{L}_{21}^2\exp\big[-2\big(\tilde{\mathsfi{L}}_{11}+\tilde{\mathsfi{L}}_{22}\big)\big] +q-p\right\}. 
\end{align}
\end{subequations}

\subsection{SSR decomposition}

 The symmetric square root of $\mathsfbi{C}$ satisfies \citep{btrd11}
\begin{equation}
\partial_t \mathsfbi{B} + \boldsymbol{u}\bcdot\bnabla\mathsfbi{B} = \mathsfbi{B}\bcdot\bnabla\boldsymbol{u}+\mathsfbi{a}\bcdot\mathsfbi{B}+\frac{1}{2\Wi}
[(\mathsfbi{B}^\top)^{-1}-f(\operatorname{tr}\mathsfbi{B})\mathsfbi{B}],
\end{equation}
where $\mathsfbi{a}$ is antisymmetric and, for $d=2$, its nonzero elements are
\begin{equation}
\mathsfi{a}_{12}=-\mathsfi{a}_{21}=\frac{\mathsfi{B}_{12}\partial_x u_x-\mathsfi{B}_{11}\partial_x u_y+\mathsfi{B}_{22}\partial_y u_x-\mathsfi{B}_{12}\partial_y u_y}{\mathsfi{B}_{11}+\mathsfi{B}_{22}}.
\end{equation}

\section{Comparison of matrix decompositions with the FENE-P model}
\label{app:fenep}

\begin{figure}
\centering
\begin{subfigure}[b]{\textwidth/2}
\includegraphics[width=\textwidth]{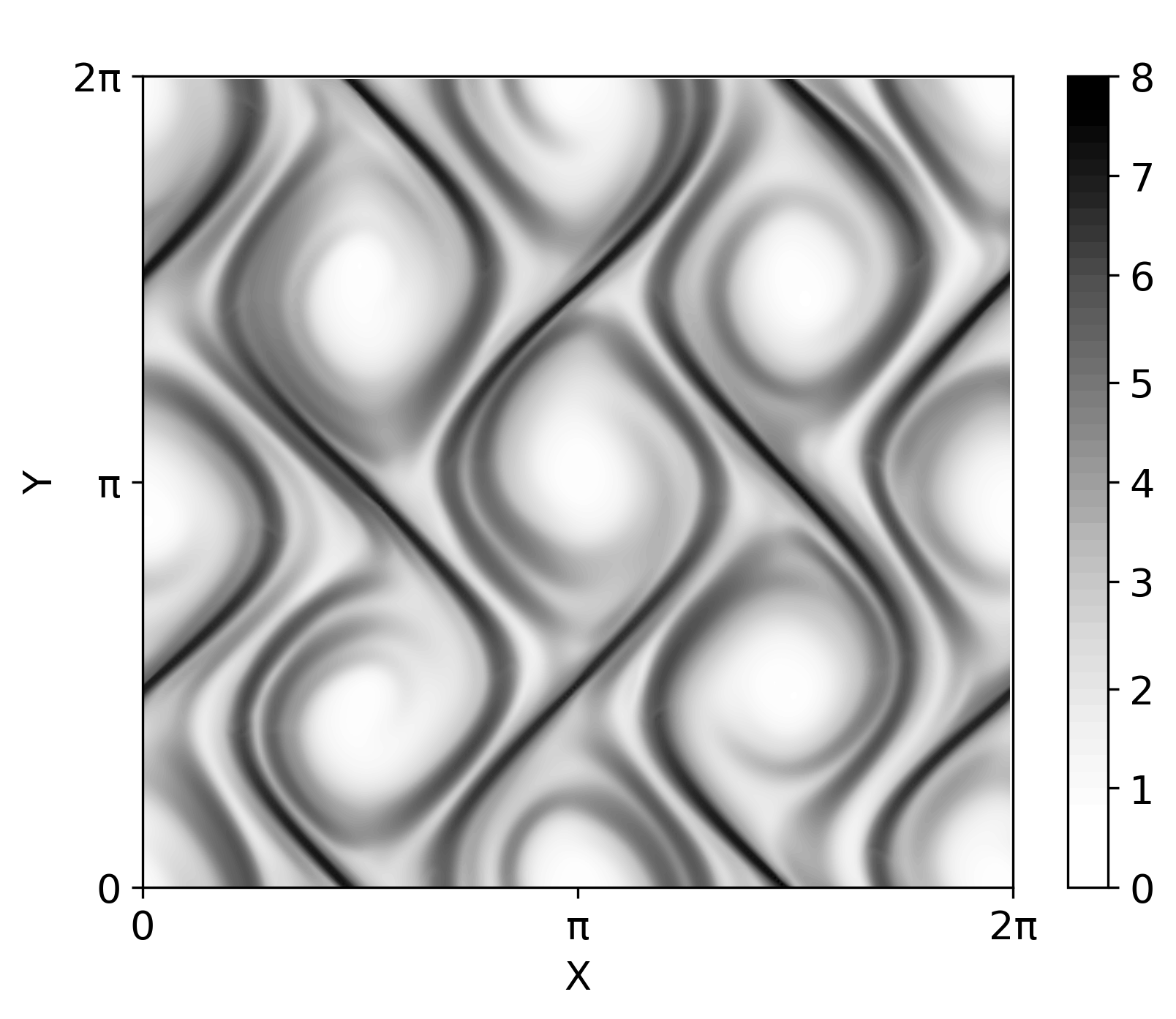}%
\label{OK}
\caption{}
\end{subfigure}%
\hfill
\begin{subfigure}[b]{\textwidth/2}
\includegraphics[width=\textwidth]{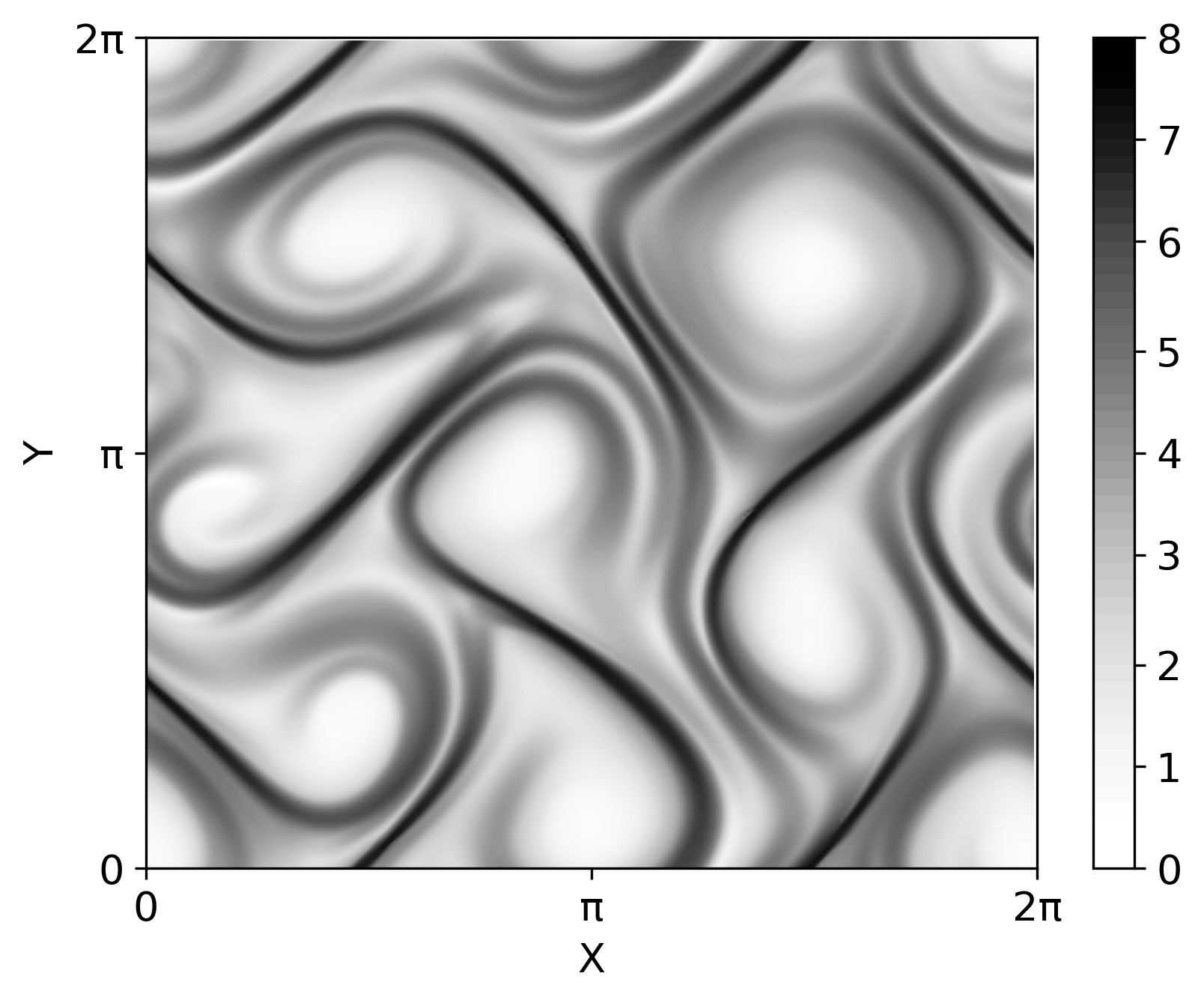}%
\caption{}
\end{subfigure}%
\caption{Representative snapshots (instantaneous) of the logarithm of $\trC$ for (\textit{a}) the Cholesky-log and (\textit{b})~the SSR decompositions. For the corresponding animations, see supplementary \href{https://math.unice.fr/~vincenzi/Movie5.mp4}{movie 5}. Both simulations use the FENE-P model.}
\label{fig:snapshots-fenep}
\end{figure}
\begin{figure}
\centering
\begin{subfigure}[b]{\textwidth/3}
\centering
\includegraphics[width=\textwidth]{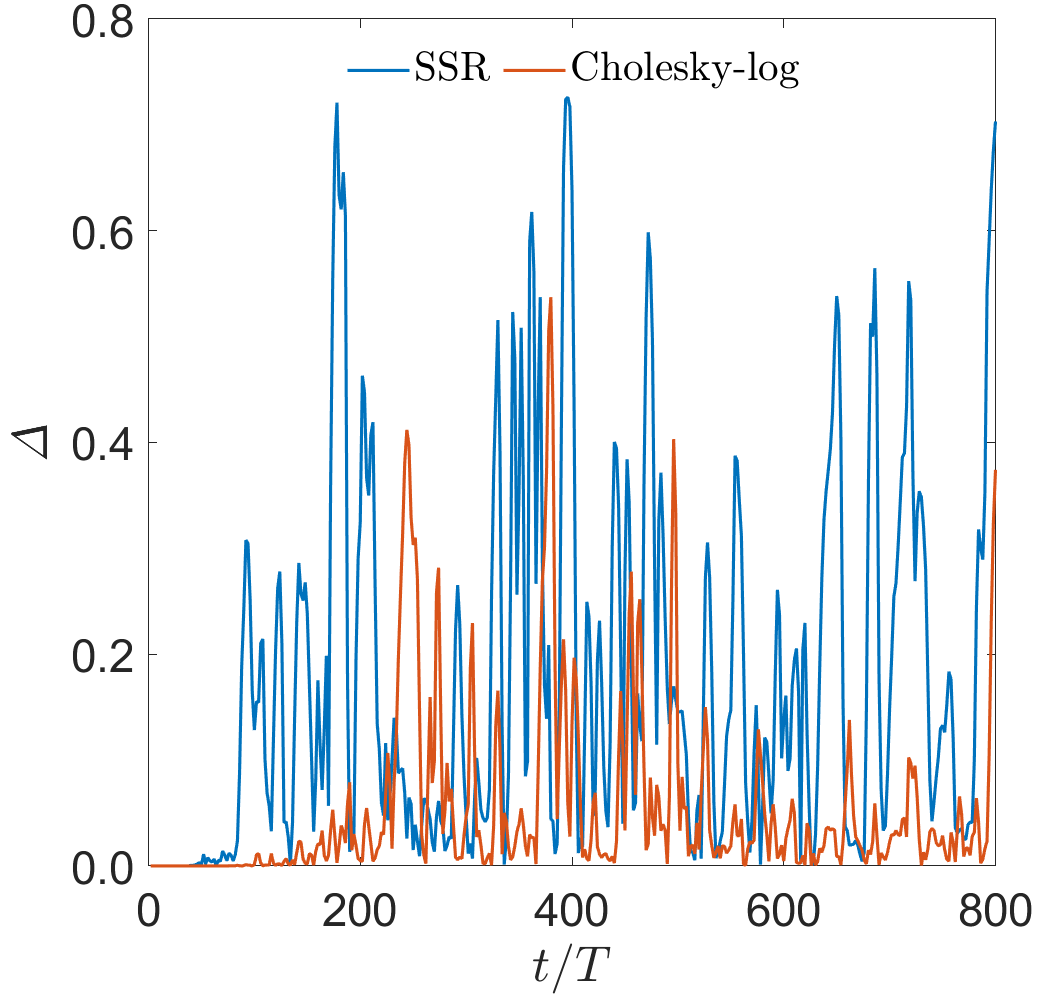}%
\caption{}
\end{subfigure}%
\begin{subfigure}[b]{\textwidth/3}
\centering
\includegraphics[width=1.0\textwidth]{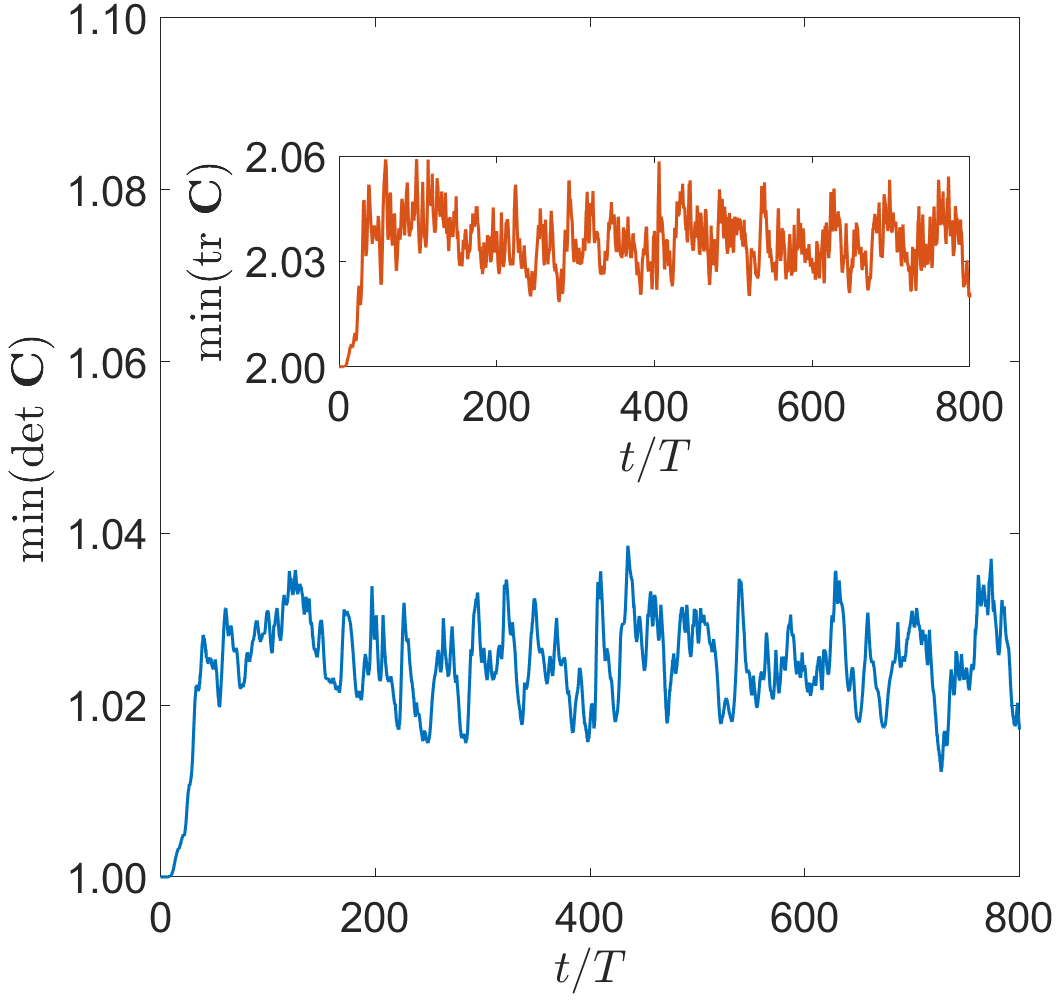}%
\caption{}
\end{subfigure}%
\hfill%
\begin{subfigure}[b]{\textwidth/3}
\centering
\includegraphics[width=0.99\textwidth]{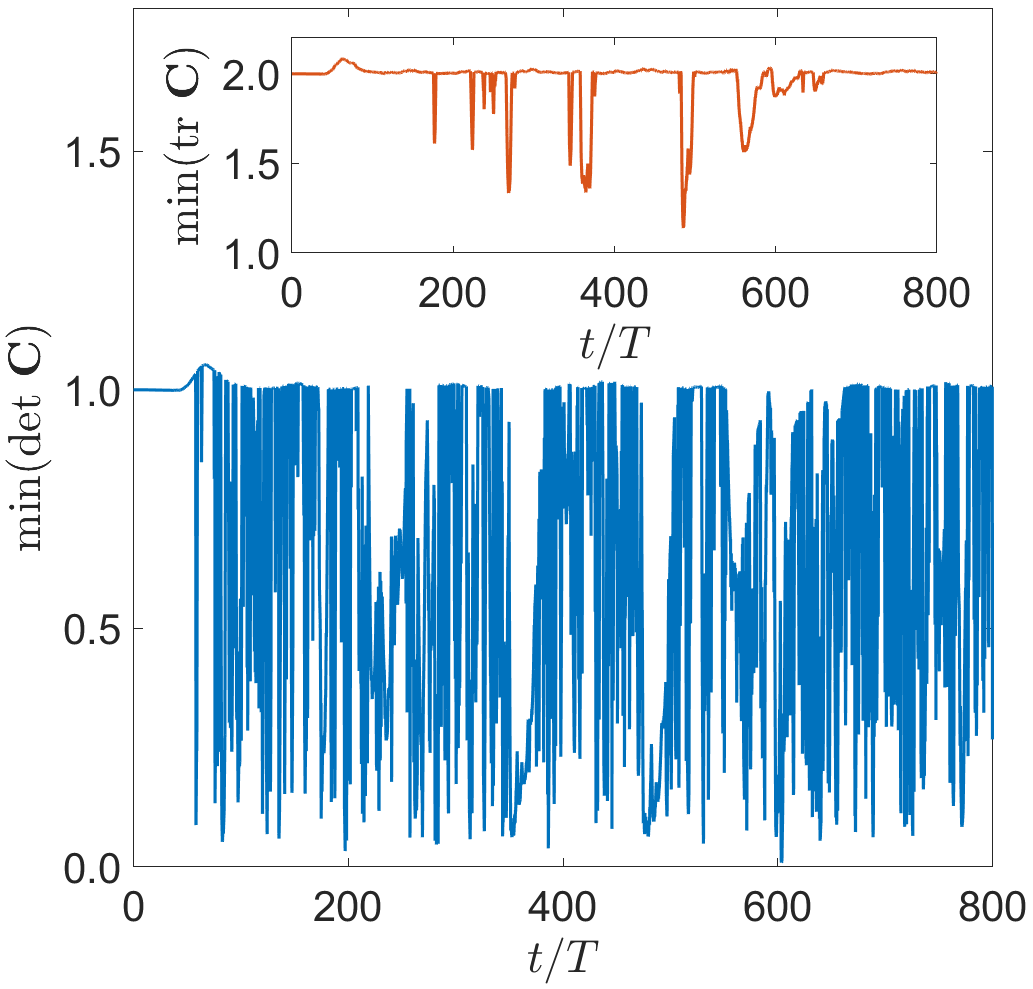}%
\caption{}
\end{subfigure}%
\caption{Time series of  (\textit{a}) $\varDelta(t)$ and the minimum of $\detC$ over the domain for (\textit{b}) the Cholesky-log and (\textit{c})~the SSR decompositions. The insets show the corresponding time series of the minimum of $\trC$ over the domain. All plots refer to simulations of the FENE-P model.}
\label{fig:det-fenep}
\end{figure}

In \S~\ref{sect:decomposition}, we compared the performances of the Cholesky-log and the SSR decompositions for the Oldroyd-B model.
Here we present an analogous comparison for the FENE-P model.

The snapshots of $\trC$ in figure~\ref{fig:snapshots-fenep} and the associated animations in supplementary \href{https://math.unice.fr/~vincenzi/Movie5.mp4}{movie 5} show that in the Cholesky-log simulation the vortical cells approximately retain the lattice structure of the forcing, whereas in the SSR simulation the vortical cells are displaced, distorted, and subject to constant changes in shape and size. 
The time series of $\varDelta(t)$ for the two decompositions (see~\eqref{eq:Delta}) confirm a greater distortion of the cellular structure in the SSR simulations (figure~\ref{fig:det-fenep}\textit{a}).

These differences parallel those observed in \S~\ref{sect:decomposition} for the Oldroyd-B model. There, we could use the bound of \cite{hl07} on $\detC$ to determine which of the two decompositions yielded the correct large-scale dynamics. Since an equivalent bound is not available for the FENE-P model, we cannot in principle draw any rigorous conclusions from the analysis of $\detC$ in this case. Nevertheless, it is interesting to note that the behaviour of $\detC$ in the FENE-P model is similar to that observed in the Oldroyd-B model: the minimum of $\detC$ over the spatial domain is persistently greater than unity in the Cholesky-log simulation, while it frequently falls below unity in the SSR simulation (figure~\ref{fig:det-fenep}\textit{b,c}). 

\bibliographystyle{jfm_arxiv}
\bibliography{polymers}

\end{document}